\documentclass[sigconf, nonacm]{acmart}

\usepackage{algorithm,algpseudocode}
\usepackage{array,multirow}
\usepackage{xspace}
\usepackage{pifont}
\usepackage{adjustbox}
\usepackage{xcolor}
\usepackage{colortbl}
\usepackage{pbox}
\usepackage{graphicx}
\usepackage{subfigure}

\def\ojoin{\setbox0=\hbox{$\bowtie$}%
	\rule[0.15ex]{.22em}{.6pt}\llap{\rule[0.9ex]{.22em}{.6pt}}}

\def\fullouterjoin{\mathbin{\ojoin\mkern-5.5mu\bowtie\mkern-5.5mu\ojoin}}

\newcommand{\CE}{\textsf{CardEst} }
\newcommand{\CEend}{\textsf{CardEst}}

\makeatletter
\newcommand*\bigcdot{\mathpalette\bigcdot@{.5}}
\newcommand*\bigcdot@[2]{\mathbin{\vcenter{\hbox{\scalebox{#2}{$\m@th#1\bullet$}}}}}
\makeatother
\newcommand{\tabincell}[2]{\begin{tabular}{@{}#1@{}}#2\end{tabular}}

\newtheorem{theorem}{Theorem}
\newtheorem{definition}{Definition}

\begin{document}
\title{BayesCard: Revitalizing Bayesian Networks for Cardinality Estimation}

\author{Ziniu Wu$^{1}$,  Amir Shaikhha$^{2}$, Rong Zhu$^{1}$, Kai Zeng$^{1}$, Yuxing Han$^{1}$, Jingren Zhou$^{1}$}

\affiliation{
\smallskip \smallskip
 \institution{\LARGE \textit{$^{1}$Alibaba Group, $^2$University of Edinburgh} \smallskip} \state{\Large \textsf{$^1\{$ziniu.wzn, red.zr, zengkai.zk, yuxing.hyx, jingren.zhou$\}$@alibaba-inc.com, $^2\{$amir.shaikhha$\}$@ed.ac.uk
}}}

\newcommand{\system}{Cardinal\xspace}
\newcommand{\cmark}{\ding{51}}%
\newcommand{\cmarkcur}{\color{blue}\ding{51}}%
\newcommand{\cmarktodo}{\color{red}\ding{51}}%
\newcommand{\xmark}{$-$}%
\newcolumntype{R}[2]{%
    >{\adjustbox{angle=#1,lap=\width-(#2)}\bgroup}%
    l%
    <{\egroup}%
}
\newcommand*\rot{\multicolumn{1}{R{90}{0em}|}}
\newcommand{\rott}{\multicolumn{1}{|R{60}{-1em}|}}

\makeatletter
\newcommand{\thickhline}{%
    \noalign {\ifnum 0=`}\fi \hrule height 1pt
    \futurelet \reserved@a \@xhline
}
\newcolumntype{"}{@{\hskip\tabcolsep\vrule width 1pt\hskip\tabcolsep}}
\makeatother

\newcommand{\firstcell}[1]{{\cellcolor{blue!25}\textbf{#1}}}
\newcommand{\secondcell}[1]{{\cellcolor{blue!15}\textit{#1}}}
\newcommand{\thirdcell}[1]{{\cellcolor{blue!6}#1}}

\begin{abstract}
Cardinality estimation (\CEend) is an essential component in query optimizers and a fundamental problem in DBMS. A desired \CE method should attain good \emph{algorithm} performance, be stable to varied \emph{data} settings, and be friendly to \emph{system} deployment. However, no existing \CE method can fulfill the three criteria at the same time. Traditional methods often have significant algorithm drawbacks such as large estimation errors. Recently proposed deep learning based methods largely improve the estimation accuracy but their performance can be greatly affected by data and often difficult for system deployment.

In this paper, we revitalize the Bayesian networks (BN) for \CE by incorporating the techniques of probabilistic programming languages. We present \textit{BayesCard}, the \emph{first} framework that inherits the advantages of BNs, i.e., high estimation accuracy and interpretability, while overcomes their drawbacks, i.e. low structure learning and inference efficiency. This makes \textit{BayesCard} a perfect candidate for commercial DBMS deployment. Our experimental results on several single-table and multi-table benchmarks indicate \textit{BayesCard}'s superiority over existing state-of-the-art \CE methods:
\textit{BayesCard} achieves comparable or better accuracy, $1$--$2$ orders of magnitude faster inference time, $1$--$3$ orders faster training time, $1$--$3$ orders smaller model size, and $1$--$2$ orders faster updates. Meanwhile, \textit{BayesCard} keeps stable performance when varying data with different settings. We also deploy \textit{BayesCard} into PostgreSQL. On the IMDB benchmark workload, it improves the end-to-end query time by $13.3\%$, which is very close to the optimal result of $14.2\%$ using an oracle of true cardinality.
\end{abstract}

\maketitle

\section{Introduction}
Cardinality estimation (\textsf{CardEst}), which aims at predicting the result size of a SQL query without its actual execution, is a longstanding and fundamental problem in DBMS. It is the core component of query optimizers~\cite{howgoodare,4,6} to produce high-quality query plans. Although a variety of \CE methods have been proposed in the last several decades, it remains to be a notoriously challenging problem in the DB community.

\smallskip
\noindent{\underline{\textbf{Status and challenges of \textsf{CardEst}.}}}
Given a table $T$ on attributes $\{T_1, \ldots, T_n\}$ and a query $Q$, \CE is equivalent to estimating the probability of tuples in $T$ satisfying $Q$.
Therefore, the core problem of \CE is how to model the distribution of $T$ to estimate the probability of $Q$. Based on existing work~\cite{wang2020ready}, we believe that an applicable \CE method should satisfy criteria from three aspects, namely \emph{A(Algorithm)}, \emph{D(Data)} and \emph{S(System)}. (\emph{A}): the \CE algorithm itself should have high estimation accuracy, fast inference (and training) time, lightweight storage cost, and efficient updating process, in order to generate high-quality query plans~\cite{zhu2020flat, perron2019learned}.
(\emph{D}): the \CE method should maintain stable performance for different data with varied distribution, attribute correlation, domain size, and number of attributes. (\emph{S}): the \CE method should be friendly for system deployment with interpretable model, predictable behaviors, reproducible results, and easy for debugging~\cite{wang2020ready}.

The simplest \CE method assumes that all attributes are mutually independent and builds a histogram on each $T_i$. Its estimation latency is low but the error is high since correlations between attributes are ignored. Another class of methods samples tuples from $T$ for \CEend. They can be inaccurate on high-dimensional data or queries with small cardinality. These traditional \CE methods have significant algorithm drawbacks and unstable performance w.r.t. varied data but friendly for system deployment. 

Recently, numerous works attempt to utilize machine learning (ML), especially deep learning (DL) techniques for \CEend. They either build supervised models mapping featurized query $Q$ to its cardinality~\cite{7, MSCN} or learn unsupervised models of $P_T$, the joint probability distribution of table $T$, to support computing the probability of any query $Q$ on $T$~\cite{zhu2020flat, deepDB, naru}. DL-based \CE methods greatly improve the estimation accuracy but often sacrifice other algorithm aspects. More importantly, their performance can be greatly affected by data and often difficult for system deployment, such as the hyper-parameter tuning and the ``black-box'' property.

Table~\ref{ADSsummary} summarizes the status of existing \CE methods according to the \emph{ADS} criteria. We can clearly see that no existing solution satisfactorily addresses this problem.

\begin{table}[t]
    \centering
    \caption{Status of \CE methods according to \emph{ADS} criteria.}
    \vspace{-1em}
    \resizebox{\columnwidth}{!}{
    \begin{tabular}{|c|c|c|c|c|c|c|c|c|c|c|c|c|c|}
    \hline
     \multirow{2}{*}{\tabincell{c}{\CE \\ Methods}} & \multicolumn{5}{c|}{Algorithm} & \multicolumn{4}{c|}{Data} & \multicolumn{4}{c|}{System}
           \\\cline{2-14}
      & \rot{Accuracy} & \rot{Latency}  & \rot{Training} & \rot{Model Size} & \rot{Updating} & \rot{Distribution} & \rot{Correlation} &\rot{Domain} & \rot{Scale} &\rot{Debug} &\rot{Interpret} &\rot{Predict} &\rot{Reproduce}  \\\hline
       \it Histogram &\xmark &\cmark &\cmark &\cmark &\cmark &\cmark &\xmark &\cmark & \xmark &\cmark &\cmark &\cmark &\cmark \\ \cline{1-14}
       \it Sampling &\xmark &\xmark &\cmark &\cmark &\cmark &\xmark &\cmark &\xmark & \xmark &\cmark &\xmark &\cmark &\xmark \\ \cline{1-14}
       \it Naru &\cmark &\xmark &\cmark &\cmark &\xmark &\xmark &\cmark &\xmark & \xmark &\xmark &\xmark &\xmark &\xmark \\ \cline{1-14}
       \it DeepDB &\cmark &\cmark &\cmark &\cmark &\xmark &\cmark &\xmark &\cmark & \xmark &\xmark &\xmark &\cmark &\cmark \\ \cline{1-14}
       \it FLAT &\cmark &\cmark &\cmark &\cmark 
       &\xmark &\cmark &\cmark &\cmark & \xmark &\xmark &\xmark &\cmark &\cmark \\ \cline{1-14}
       \it MSCN &\xmark &\cmark &\xmark &\cmark
       &\xmark &\cmark &\cmark &\cmark & \xmark &\xmark &\xmark &\xmark &\cmark \\ \cline{1-14}
       \it BN &\cmark &\xmark &\xmark &\cmark
       &\cmark &\cmark &\cmark &\cmark & \cmark &\cmark &\cmark &\cmark &\cmark \\ \cline{1-14}
       \textbf{\textit{BayesCard}} &\cmark &\cmark &\cmark &\cmark &\cmark &\cmark &\cmark &\cmark & \cmark &\cmark &\cmark &\cmark &\cmark \\ \cline{1-14}
    \end{tabular}}
    \vspace{-1.5em}
    \label{ADSsummary}
\end{table}

\smallskip
\noindent{\underline{\textbf{Our motivation.}}}
Recently, a classical method Bayesian networks (BNs) have re-attracted numerous attentions in the ML community to overcome the drawbacks of deep models~\cite{zhu2020efficient,lee2019scaling,ye2020optimizing}, and they are naturally suitable for \CE~\cite{2001SigmodGreedy, tzoumas2013vldb, dasfaa2019}. In comparison with other methods, BNs have significant advantages in terms of the \emph{ADS} criteria.
First, from the algorithm perspective, BNs are very compact and easy to update.
Second, BNs reflect the intrinsic causal relations between attributes, which are robust to the data changes. Thus, they tend to maintain stable performance as the data varies in terms of correlation, distribution, and etc.
Third, BNs are \emph{interpretable}, easy to predict, maintain, validate and improve with expert knowledge, thus friendly for system deployment.
These attractive models have been proposed decades ago~\cite{2001SigmodGreedy, tzoumas2013vldb}, but the BNs' NP-hard model construction process and intractable probability inference algorithm make them impractical for DBMS.

\emph{In summary, as long as we can overcome the inefficiency of model construction and probability inference of BNs, we can obtain a desirable method for \CE satisfying the ADS criteria simultaneously.}

\smallskip
\noindent{\underline{\textbf{Our contributions.}}}
In this paper, we try to resolve the \CE challenges by revitalizing BNs with new equipments. We propose \textit{BayesCard}, a unified Bayesian framework for \CEend. The key idea of \textit{BayesCard} is to build an ensemble of BNs to model the distributions of tables in a database, and use the constructed BNs to estimate the cardinality of any query. \textit{BayesCard} incorporates the recent advances in probabilistic programming languages (PPLs) for building BNs~\cite{edward,pyro,InferNET18,schreiber2018pomegranate,ankan2015pgmpy, pymc}.
PPLs allow for a declarative specification of probabilistic models, within which each variable is defined as a probability distribution influenced by others. 
Based on PPLs, we can easily define BNs to support various structure learning, parameter learning, and inference algorithms. Therefore, \textit{BayesCard} provides a user-friendly framework of building different BNs suitable for various data and system settings.

The key techniques of \textit{BayesCard} overcome the deficiency of existing BNs.
First, based on PPLs, \textit{BayesCard} designs the \emph{progressive sampling} and \emph{compiled variable elimination} probability inference algorithms, which significantly accelerate the traditional BN's inference process. Moreover, \textit{BayesCard} adapts its inference algorithms to efficiently handle multi-table join queries. Second, \textit{BayesCard} designs an efficient model construction algorithm for building an ensemble of BNs. Furthermore, using PPLs, \textit{BayesCard} can pre-specify constraints on the learned BN structure with prior knowledge to speed up the structure learning process. An accurate and lightweight BN structure could be obtained efficiently. 


By our benchmark evaluation results, in comparison with DL-based \CE methods, \textit{BayesCard} achieves comparable or better accuracy, $1$--$2$ orders of magnitude lower inference latency (near histogram) and update time, and $1$--$3$ orders faster training time and smaller model size.
Meanwhile, \textit{BayesCard} keeps stable performance when varying data with different settings.
We also integrate \textit{BayesCard} into PostgreSQL. On the benchmark workload, it improves the end-to-end query time by $13.3\%$, which is very close to the optimal result of $14.2\%$ using the true cardinality. 

In summary, the main contributions of this paper are as follows:

\indent $\bullet$
We analyze the existing \CE methods in terms of the \emph{ADS} criteria to evaluate a good and practical \CE method. (Section~\ref{sect2})

\indent $\bullet$
We propose \textit{BayesCard}, a general framework that unifies the efforts behind PPLs for constructing BNs for \textsf{CardEst}. (Section~\ref{sect3})

\indent $\bullet$
We develop algorithms and techniques in \textit{BayesCard} using PPLs to improve inference latency and reduce the model construction cost, which help \textit{BayesCard} attain the desired properties of \CE methods.
(Section~\ref{sect4} and~\ref{sect5})

\indent $\bullet$ We conduct extensive experiments on benchmarks and integrate \textit{BayesCard} into real-world system to demonstrate its superiority from \textit{ADS} criteria. (Section~\ref{sect6})

\section{Problem definition and Analysis}
\label{sect2}
In this section, we first formally define the \CE problem from both database and statistical perspectives and then exhaustively examine the existing traditional methods and state-of-the-art DL-based methods for \CE from the \emph{ADS} criteria. 


    
    
    
    
    
    



\smallskip
\noindent\underline{\textbf{\CE problem.}}
Let $T$ be a table with $n$ attributes $T_1, \cdots, T_n$.
For each $1 \leq i \leq n$, let $D_i$ denote the domain (all unique values) of attribute $T_i$. Any selection query $Q$ on $T$ can be represented in a canonical form\footnote{Handling pattern matching queries or string predicates (e.g., ``LIKE'' queries) require extensions (such as q-grams~\cite{chaudhuri2004selectivity}), which we do not consider in this paper.} as $Q = \{T_1 \in R_Q(T_1) \wedge T_2 \in R_Q(T_2) \wedge \cdots \wedge T_n \in R_Q(T_n)\}$, where $R_Q(T_i) \subseteq D(T_i)$ is the region specified by $Q$ over attribute $T_i$. Without loss of generality, we have
$R_Q(T_i) = D(T_i)$ if $Q$ has no constraint on attribute $T_i$.


Let $C_Q$ denote the cardinality, i.e., the number of tuples in $T$ satisfying query $Q$. From a statistical perspective, we can also regard all tuples in $T$ as points sampled according to the joint distribution $P_T = P_T(T_1, T_2, \dots, T_n)$ of all attributes. Let $P_T(Q) = P_T(T_1 \in R_Q(T_1), T_2 \in R_Q(T_2), \cdots , T_n \in R_Q(T_n)$ be the probability specified by the region of $Q$. Then, we have $C_Q = P_T(Q) \cdot |T|$. Thus, the \CE problem can essentially be reduced to model the probability density function (PDF) $P_T$ of table $T$. In this paper, we focus on data-driven \CE methods, which try to model $P_T$ directly. For query-driven \CE methods,
they implicitly model $P_T$ by building functions mapping $Q$ to $P_T(Q)$.

\smallskip
\noindent\underline{\textbf{Existing \CE Methods.}} We review the two traditional methods widely used by commercial DBMS and four state-of-the-art (SOTA) DL-based methods.

\textit{1). Histogram}~\cite{10} method assumes all attributes in $T$ are independent, and thus $P_T$ can be estimated as the $\prod_{i=1}^n P_T(T_i)$. 

\textit{2). Sampling} is a model-free method, which fetches tuples from $T$ on-the-fly to estimate the probability of $Q$ on the samples. 

\textit{3). Naru}~\cite{naru}, based on deep auto-regression models (DAR)~\cite{made}, factorizes $P_T$ as $P_T(T_1) * \prod_{i=2}^{n} P_T(T_i|T_1,\ldots,T_{n-1})$ and approximate each conditional PDF by a deep neural network (DNN). 

\textit{4). DeepDB}~\cite{deepDB}, based on sum-product networks (SPN)~\cite{SPN}, approximates $P_T$ by recursively decomposing it into local and simpler PDFs. Specifically, the tree-structured SPN contains sum node to split $P_T$ to multiple $P_{T'}$ on tuple subset $T' \subseteq T$, product node to decompose $P_{T'}$ to $P_{T'}(T_i) \cdot P_{T'}(T_j)$ if attributes $T_i$ and $T_j$ are independent and leaf node if $P_{T}$ is a univariate PDF.

\textit{5). FLAT}~\cite{zhu2020flat}, based on factorized-split-sum-product networks (FSPN)~\cite{wu2020fspn}, improves over SPN by
adaptively decomposing $P_T$ according to the attribute dependence level. It adds the factorize node to split $P_T$ as $P_T(W) \cdot P_T(H | W)$ where $H$ and $W$ are highly and weakly correlated attributes in $T$. $P_T(W)$ is modeled in the same way as SPN. $P_T(H | W)$ is decomposed into small PDFs by the split nodes until $H$ is locally independent of $W$. Then, the multi-leaf node is used to model the multivariate PDF $P_T(H)$ directly.

\textit{6). MSCN}~\cite{MSCN}, is a query-driven method, which uses the set-convolutional DNN to learn the mapping functions between the input query $Q$ and its probability $P_T(Q)$. 

\smallskip
\noindent\underline{\textbf{Analysis Results.}} We elaborate the \emph{ADS} criteria for \CE problem and analyze the aforementioned methods in details. The results are summarized in Table~\ref{ADSsummary}.  

\noindent\textit{$\bullet$ \textbf{Algorithm.}} 
From the algorithm's perspective, we consider five important metrics that are widely used in existing work~\cite{deepDB, zhu2020flat} to evaluate the performance of \CE methods. 

$\bigcdot$
\emph{Estimation accuracy} is one of the priorities for \CE since inaccurate estimation typically leads to sub-optimal and slow query plan~\cite{howgoodare}. Unfortunately, the traditional methods frequently incur poor estimations: \emph{Histogram} can cause large estimation error in presence of attributes correlations and \emph{Sampling} may be inaccurate on high-dimensional data with limited sampling size.
Query-driven methods, such as \emph{MSCN}, also have poor accuracy if the target query does not follow the same distribution of the query workload that the model is trained on. By existing evaluations~\cite{naru, deepDB, zhu2020flat}, DL-based \CE methods can produce accurate results. 

$\bigcdot$
\emph{Inference latency} is crucial since \CE method needs to be executed numerous times in query optimization~\cite{3,6}. As a result, slow latency may degrade the end-to-end query time on plan generation and execution. 
The inference latency of \emph{Naru} is high because of its large underlying DNN models and repetitive sampling process. \emph{Sampling} is also not efficient when the sample size is large. 

$\bigcdot$
\emph{Training cost} refers to \CE model construction time for a given database.
Query-driven based methods, such as \emph{MSCN}, are in general slow for training, since an enormous amount of queries need to be executed to learn the models.

$\bigcdot$
\emph{Model size} is related to the storage cost of models. In nowadays DBMS, the space costs of all these \CE methods are affordable. 

$\bigcdot$
\emph{Update time} is also important since table data frequently changes. Traditional methods are easy to update while no existing DL-based method can keep up with the fast data updates~\cite{wang2020ready}. 

\smallskip
\noindent\textit{$\bullet$ \textbf{Data.}} 
Generally, a DBMS will process various data with different settings. Therefore, we analyze whether the \CE methods have a stable performance on four typical variations of data settings, namely data \textit{distribution}, attribute \textit{correlation}, attribute \textit{domain} size, and the number of attributes (\textit{scale}).

For traditional methods, \emph{Histogram}'s estimation error grows exponentially when data are highly correlated. \emph{Sampling}'s accuracy degrades on high-dimensional data with larger domain size and more attributes. In addition, for highly skewed data, the fetched samples tend to miss the query ranges with small probability, which also degrades its accuracy.

For DL-based methods, the poor performance stability of \emph{Naru}, \emph{DeepDB} and \emph{MSCN} is demonstrated in a recent benchmark study~\cite{wang2020ready}.
In a nutshell, their accuracy decreases while inference and training cost increases with more attributes. \emph{Naru} is also sensitive to data distribution and domain size since skewed or large PDF is more difficult to model. \emph{DeepDB} has the intrinsic drawback that tends to generate large and inaccurate SPNs on highly correlated attributes~\cite{expsSPN}. \emph{FLAT} overcomes the drawback of \emph{DeepDB} but its performance also degrades severely with more attributes.

\smallskip
\noindent\textit{$\bullet$ \textbf{System.}} 
The \CE method should satisfy the following properties for friendly system deployment~\cite{wang2020ready}.

$\bigcdot$
\emph{Debuggability} and easy to tune are crucial to the DB experts. The DL-based methods with ``black-box'' components may fail silently and contain high risks of missing a bug~\cite{wang2020ready}.

$\bigcdot$
\emph{Interpretability} is necessary when system developers would like to explain and validate the learned component, which is not satisfied by the DL-based methods~\cite{interpretation}.

$\bigcdot$
\emph{Predictability} is important since the system developers would like to predict the performance before actual deployment. As \emph{Naru} and \emph{MSCN} contain DNNs with illogical behaviors~\cite{wang2020ready}, their performance is hard to predict.

$\bigcdot$
\emph{Reproducibility} is necessary to locate system issues. As \emph{Sampling} and \emph{Naru} involve stochastic processes, their results cannot be reproduced by estimating the same query one more time.

\smallskip
\noindent\underline{\textbf{Summary.}} 
From Table~\ref{ADSsummary}, we observe that \emph{no} existing \CE method is satisfactory in all criteria. Our detailed experimental evaluation in Section~7 also verifies this observation.
Therefore, we design a new \CE framework \emph{BayesCard} that successfully satisfies all criteria for the first time.

\begin{figure*}[t]
  \centering
  \includegraphics[width=17.5cm]{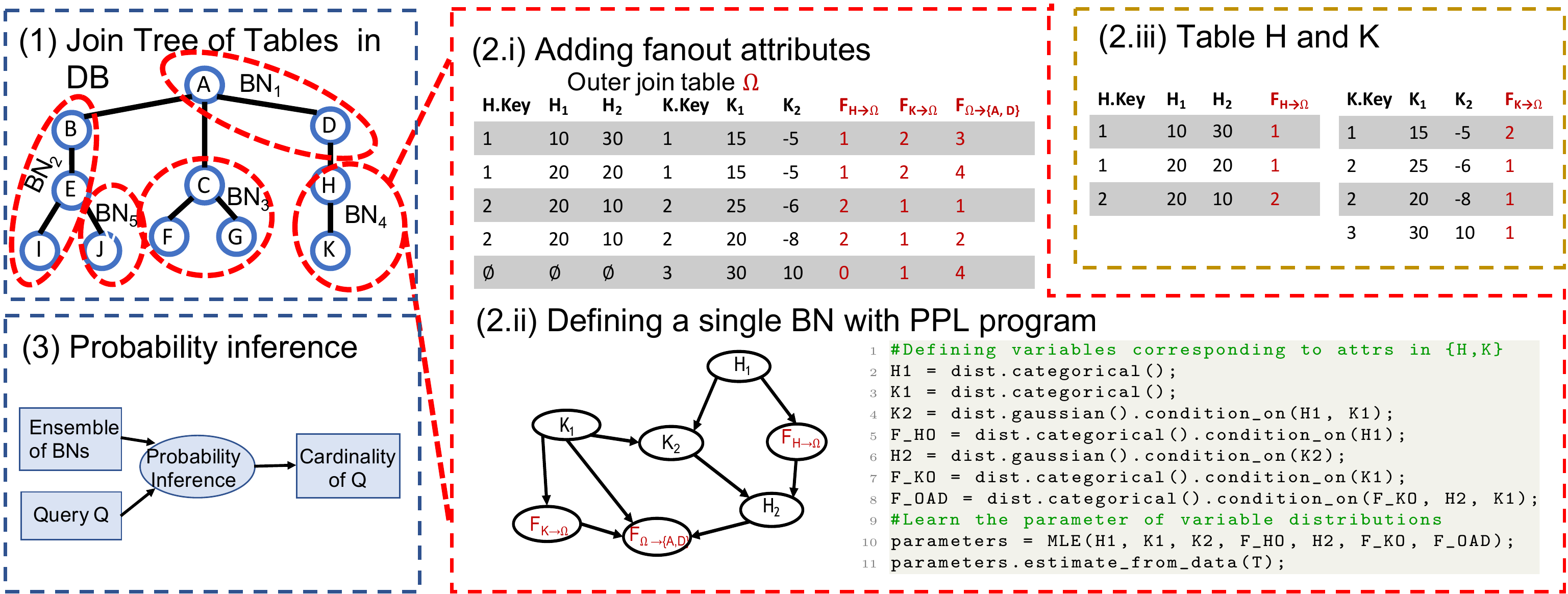}
  \caption{An example workflow of \textit{BayesCard}. }
  \label{fig_model}
\end{figure*}

\section{BayesCard Overview}
\label{sect3}

In this section, we briefly review the background knowledge on BN and PPL in Section~\ref{sect3.1}, which are the foundations of \textit{BayesCard}. Then we overview our new framework \textit{BayesCard} for \CE in Section~\ref{sect3.2}.

\subsection{Background Knowledge}
\label{sect3.1}
\noindent\underline{\textbf{Bayesian networks}} specifies a probability distribution $P_T$ of table $T$, whose attributes form a directed acyclic graph (DAG), such as Image (2.ii) in Figure~\ref{fig_model}. Each node of the DAG corresponds to an attribute and each edge defines the causal dependency between two nodes. An attribute is dependent on its parents (the source nodes with edges directing to this attribute) and conditionally independent of all other attributes given its parents~\cite{PGM}. Thus, the $P_T$ can be compactly represented as $P_T(T_1, \cdots, T_n) = \prod_{i=1}^n P_T(T_i|Par(T_i))$, where $Par(T_i)$ denotes the set of parents of $T_i$ in the defined DAG. 

\smallskip
\noindent\underline{\textbf{Probabilistic programming languages}} 
are general-purpose programming paradigm to specify probabilistic models and perform inference on the models automatically. Unlike in traditional programming languages (TPLs), each variable in PPLs is defined as a probability distribution, whose value can condition on a set of other variables. The compilers of PPLs are optimized to efficiently learn parameters of variable distribution and sample from these distributions.
PPLs have been applied to various ML domains, such as computer vision~\cite{kulkarni2015picture}, with remarkable performance.

To define a BN, for each attribute $T_i$, the PPLs can define a variable whose distribution is conditioned on variables in $Par(T_i)$. For example, the first seven lines in the PPL program on the right side of Image (2.ii) in Figure~\ref{fig_model} sufficiently defines the BN on the left as seven variables. 
PPLs have the following properties.
First, PPLs can define variables of any general distribution, including tabular and continuous distributions, which helps to build BNs with continuous attributes. Whereas, existing BNs for \CE problems~\cite{2001SigmodGreedy, dasfaa2019, tzoumas2013vldb} only support discrete variables.
Second, PPLs can efficiently learn the parameters using maximum likelihood estimation (MLE)~\cite{InferNET18}; e.g. the parameters of the example BN in Image (2.ii) can be derived by simply executing the last two lines of code. 
Third, PPLs~\cite{pomegranate} also incorporates several main-stream algorithms for learning the BNs' structure, which captures the causal pattern of attributes in the data. The structure learning procedure of PPLs supports pre-specifying sub-structures.
Forth, PPLs can efficiently generate samples from the distribution of each variable.

\vspace{1em}
\subsection{BayesCard framework}
\label{sect3.2}

In this paper, we propose \textit{BayesCard}, a framework for \CEend. 
The key idea of \textit{BayesCard} is to build an ensemble of BNs to model the distributions of tables in a database and use the constructed BNs to estimate the cardinality of any query. This framework, including model construction and probability inference of BNs, is implemented using PPLs in order to leverage its compiler and execution advantages of presenting probability distribution.


Specifically, the inputs of \textit{BayesCard} are a DB $\mathcal{D}$ containing $n$ tables and its join schema $\mathcal{J}$. 
Following prior work's assumption~\cite{zhu2020flat, NeuroCard, Sampling}, \textit{BayesCard} only considers the join schema to be a tree, i.e. without self joins or cyclic joins.
In the join tree $\mathcal{J}$, each node represents a table and each edge represents a join relation between two tables.
For example, Figure~\ref{fig_model}-(1) illustrates a DB with 11 tables and the join tree schema on the tables.

Given $\mathcal{D}$ and $\mathcal{J}$, \textit{BayesCard} constructs an ensemble of $m$ BNs. 
Each BN models the joint distribution of a subset of connected tables in $\mathcal{J}$.
For example in Figure~\ref{fig_model}-(1), \textit{BayesCard} builds 5 BNs ($BN_1, \ldots, BN_5$ in the red circles) to characterize the distributions of tables in the DB, where $BN_4$ is built to represent the joint distribution of tables $H$ and $K$.

To accurately model the joint distribution of multiple tables $\mathcal{T}$, \textit{BayesCard} uses the \emph{fanout} method as in prior works~\cite{deepDB, zhu2020flat, NeuroCard}, by creating a BN on the full outer join results of $\mathcal{T}$, along with additional fanout attributes. For example, as shown in Figure~\ref{fig_model}-(2.i), $BN_4$ models $\Omega$, the full outer join of $H$ and $K$ (shown in Figure~\ref{fig_model}-(2.iii)), along with the added fanout attributes:
$F_{H\xrightarrow{}\Omega}$, indicating how many tuples in $\Omega$ does a particular tuple in $H$ fanouts to; $F_{K\xrightarrow{}\Omega}$, indicating how many tuples in $\Omega$ does a particular tuple in $K$ fanouts to; $F_{\Omega \xrightarrow{} \{A,D\}}$, indicating how many tuples in the outer join table $\Omega \fullouterjoin A \fullouterjoin D$ does a particular tuple in $\Omega$ fanouts to. 

Each BN can be represented as a PPL program, such as $BN_4$ in Figure~\ref{fig_model}-(2.ii). The probability $P_{\mathcal{T}}(Q)$ of any query $Q$ on a subset of tables $\mathcal{T}$ can be estimated based on the combination of multiple BNs containing tables covered in $\mathcal{T}$. The process of estimating the probability of a given query $P_{\mathcal{T}}(Q)$ is called probability inference.

\smallskip
\noindent \underline{\textbf{Challenges.}} Existing PPLs are not optimized for \CE tasks in terms of probability inference and model construction, which are all addressed and optimized in \textit{BayesCard}.

\noindent \textbf{Probability inference.} 
After the PPL program is successfully declared to represent a BN, existing PPLs do not support using this program for efficient probability inference, which is the key to \CE problem. Therefore, \textit{BayesCard} tailors existing PPLs and designs two efficient inference algorithms. Using PPLs' extremely efficient sampling process, \textit{BayesCard} proposes the \emph{progressive sampling} algorithm, which guarantees to run in linear time complexity for estimating any query (Section~\ref{sect4.1}). In addition, \textit{BayesCard} invents \emph{compiled variable elimination} to further accelerate the inference algorithm (Section~\ref{sect4.2}). Furthermore, \textit{BayesCard} adapts its inference algorithms for the \emph{fanout} method to efficiently combine results from multiple BNs to estimate the probability of join queries (Section~\ref{sect4.3}).


\noindent \textbf{Model construction.} 
A database generally contains multiple tables and deciding which ensemble of BNs corresponding to the partition of tables to learn significantly affects the \CE accuracy and efficiency. Therefore, \textit{BayesCard} designs the ensemble construction algorithm to explore the optimal partition of all tables in the DB and optimizes the \CE quality (Section~\ref{sect5.1}).
Furthermore, Existing PPLs do not explore how to accelerates the structure learning algorithms in DB scenarios. \textit{BayesCard} tailors and speeds up these algorithms by exploring and exploiting functional dependencies and other user-defined expert knowledge (Section~\ref{sect5.2}).

\section{Probability Inference in BayesCard}
\label{sect4}

In this section, we address the \emph{probability inference} in \textit{BayesCard}. Specifically, we first propose two novel inference algorithms based on PPLs for a single BN model, namely \emph{progressive sampling} (Section~\ref{sect4.1}), which guarantees to return an approximate probability estimation in linear time, and \emph{complied variable elimination} (Section~\ref{sect4.2}), which returns the exact probability with two orders of magnitude acceleration. Next, we present how to extend these two algorithms on multiple BNs to support join queries (Section~\ref{sect4.3}).

\subsection{Progressive sampling}
\label{sect4.1}

\begin{algorithm}[t]
\small
\caption{Progressive Sampling Inference Algorithm}
\label{prog_samp_algo}
\begin{flushleft}
\textbf{Input}: a table $T$ with $n$ attributes, a query $Q$ with region $R_{Q}$ and a PPL program defining the BN on $P_T$
\end{flushleft}
\begin{algorithmic}[1]
\State Align the attributes in topological order $T_1, \ldots, T_n$
\State $p \gets 1$, $S \gets [0]_{k \times n}$, an $k \times n$ dimension matrix of samples
\For{$i \in \{1, \ldots, n\}$}
    \State Take $S[Par(T_i)]$, the columns in $S$ corresponding to attributes in $Par(T_i)$
    \State $\hat{P_i}(T_i) \gets \frac{1}{k} \sum_{d \in S[Par(T_i)]} P_T(T_i|d)$
    \State $p \gets p * \hat{P_i}(T_i \in R_Q(T_i))$
    \State Define a PPL variable $P'_i$ by normalizing $\hat{P_i}(t_i|t_i \in R_Q(T_i))$
    \State $S[i] \gets $ $k$ points sampled from $P'_i$
    \EndFor
\State \textbf{return}  $p$
\end{algorithmic}
\end{algorithm}

We define the inference procedure of a simple case, where we have a query $Q$ on tables $T$ in a DB and a single BN that exactly models $P_T$ on the full outer join of tables $T$. In this case, estimating the cardinality of $Q$, $P_T(Q)$ can be derived directly on this BN.
As defined in Section~\ref{sect2}, a query $Q$ takes the form of $\{T_1 \in R_Q(T_1) \wedge T_2 \in R_Q(T_2) \wedge \cdots \wedge T_n \in R_Q(T_n)\}$, where $R_Q$ is the region defined by $Q$ over attributes in $T$. 

Thus, we can represent the probability of $Q$ as:
$P_T(Q) = \prod_{i=1}^n P_T \break (T_i \in R_Q(T_i)|Par(T_i) \in R_Q(Par(T_i))) = \prod_{i=1}^n P_i$, where $R_Q(Par(T_i))$ denotes the query region over the set of parent attributes $Par(T_i)$ and we can denote each term as $P_i$, for simplicity. Therefore, to compute $P_T(Q)$, we only need to compute or estimate each $P_i$.

In PPLs, accessing the probability $P_T(T_i|s)$ for each fixed value assignment $s \in R_Q(Par(T_i))$ takes constant time complexity. However, computing $P_i$ is generally intractable, as there can be exponential or infinite number of unique values in $R_Q(Par(T_i))$. Specifically, for large BNs with complex structures, the PPLs' existing inference algorithms can not have an efficiency guarantee, which is required for \CE in practical DBMS. Therefore, \textit{BayesCard} designs the \emph{progressive sampling} inference algorithm, which uses the Monte Carlo approximation of $P_i$ based on a sample $S$ of $R_Q(Par(T_i))$ to ensure the computation efficiency, i.e., $P_i \approx \frac{1}{|S|} \sum_{s \in S} P_T(R_Q(T_i)|s)$. 

The default sampling procedure in PPLs only supports sampling values from a variable's domain, which are not like to fail in the query range $R_Q$. Naively using this sampling algorithm will result in enormous ineffective points. Therefore, we can leverage the learned model, create variables to materialize the distribution $P(Par(T_i)| Par(T_i) \in R_Q(Par(T_i)))$, and progressively sample points from $R_Q(Par(T_i))$ accordingly, which greatly improves the sample effectiveness.


\smallskip
\noindent \underline{\textbf{Algorithm description.}} We present the details in Algorithm~\ref{prog_samp_algo}. Specifically, we first align the attributes from $T$ in topological order as $T_1, \ldots, T_n$, where $T_1$ is the root of the BN's DAG structure (line 1). We can directly obtain from the PPL $P_T(T_1)$ as it does not depend on any other attribute, and compute $P_1 = P_T(R_Q(T_1))$. Then, we can define a new variable in PPLs to represent the distribution $P_T(t_1|t_1 \in R_Q(T_1))$ and generate sample $S_1$ of $R_Q(T_1)$ from this variable. Next, for each of the rest attributes $T_i$, the samples of its parents $Par(T_i)$ must have already been generated because the attributes are aligned in topological order (line 5). We can derive a new distribution $\hat{P_i}$ approximating $P_T(T_i | R_Q(Par(T_i)))$ using these samples (line 6). This distribution $\hat{P_i}$ will be used to estimate $P_i$ (line 7) and generate samples from $R_Q(T_i)$ (line 8). At last, after we achieve the estimated value for each $P_i$, $P_T(Q)$ can be computed as their product (line~10).

\smallskip
\noindent \underline{\textbf{Analysis.}}
Sampling $|S|$ points and evaluating the probability with each fixed point takes $O(|S|)$ time complexity to approximate each $P_i$. Thereafter, time complexity of \emph{progressive sampling} on BN with any structure is guaranteed to be $O(|S|*n)$. 
This inference algorithm is very efficient because generally, a small sample $S$ would suffice to make a very accurate estimation and the sampling process is extremely efficient in PPL. The progressive sampling algorithm in PPL resembles the one in the DAR model, proposed by Naru~\cite{naru}. Our method is different from theirs in the following aspects: 1) Efficient sampling is naturally supported in PPL for various continuous distributions, whereas the sampling procedure in DAR is post-equipped for categorical distributions only. 2) The progressive sampling in \textit{BayesCard} estimates each $P_i$ using sample $S$ during the sampling process, whereas in DAR, the samples $S$ are used to directly compute the $P_T$, which is less effective.

\smallskip
\noindent \underline{\textbf{Graph reduction optimization.}} To further accelerate the \emph{progressive sampling} algorithm, \textit{BayesCard} proposes the graph reduction optimization, which significantly speeds up the inference latency for datasets with large amount of attributes.

\noindent \textbf{Main idea.} In fact, the \emph{progressive sampling} algorithm involves a large amount of redundant computation.  For example, for an attribute $T_i$, which is not constrained by predicates in $Q$, i.e. $R_Q(T_i) = D(T_i)$, the estimation of $P_i$ should equal to $1$. If all the decedents $T_j$ of $T_i$ are not constrained in $Q$, there is no need to sample $T_i$ since each $P_j$ should equal to $1$ regardless of the samples. Therefore, we can reduce the larger BN model to a much smaller one by removing these redundant attributes, and perform probability inference on it without affecting the estimation accuracy.

\noindent \textbf{Formulation.} First, we make the following rigorous definition of reduced graph $G'$. Intuitively, $G'$ only contains all constrained attributes in the query and other necessary attributes to connect them to form a minimal BN. An example of a reduced graph can be found in Figure~\ref{fig_RG}.

\begin{definition}
Given a BN representing a table $T$ with attributes $V$ = \{$T_1, \cdots T_n$\}, its defined DAG $G = (V, E)$, and a query $Q=(T'_1=t'_1 \wedge \cdots \wedge T'_k=t'_k)$ where $T'_i \in V$. We define the reduced graph $G' = (V', E')$ to be a sub-graph of $G$ where $V'$ equals $\bigcup_{1\leq i \leq k} Ancestor(T'_i)$, and $E'$ equals all edges in $E$ with both endpoints in $V'$. $Ancestor(T'_i)$ includes all parent nodes of $T'_i$ and their parent nodes recursively.
\end{definition}

Based on this definition, we can reduce the original BN model (i.e. PPL program with variables $V$) into a much smaller one (i.e. PPL program with variable $V'$), and perform inference on it.  The correctness of the graph reduction optimization is stated in Theorem~1. Due to space limits, we put the proof of all theorems in the Appendix~A of the accompanied technical report~\cite{wu2020bayescard}. 

\begin{theorem}
\label{thm_rg}
Given a BN $B$ defining $G$, a query $Q$ and the reduced BN $B'$ defining $G'$ on $Q$, computing $P_T(Q)$ on $B'$ is equivalent to computing $P_T(Q)$ on $B$.
\end{theorem}

\subsection{Compiled variable elimination}
\label{sect4.2}

\begin{figure}[t]
  \centering
  \includegraphics[width=8.5cm]{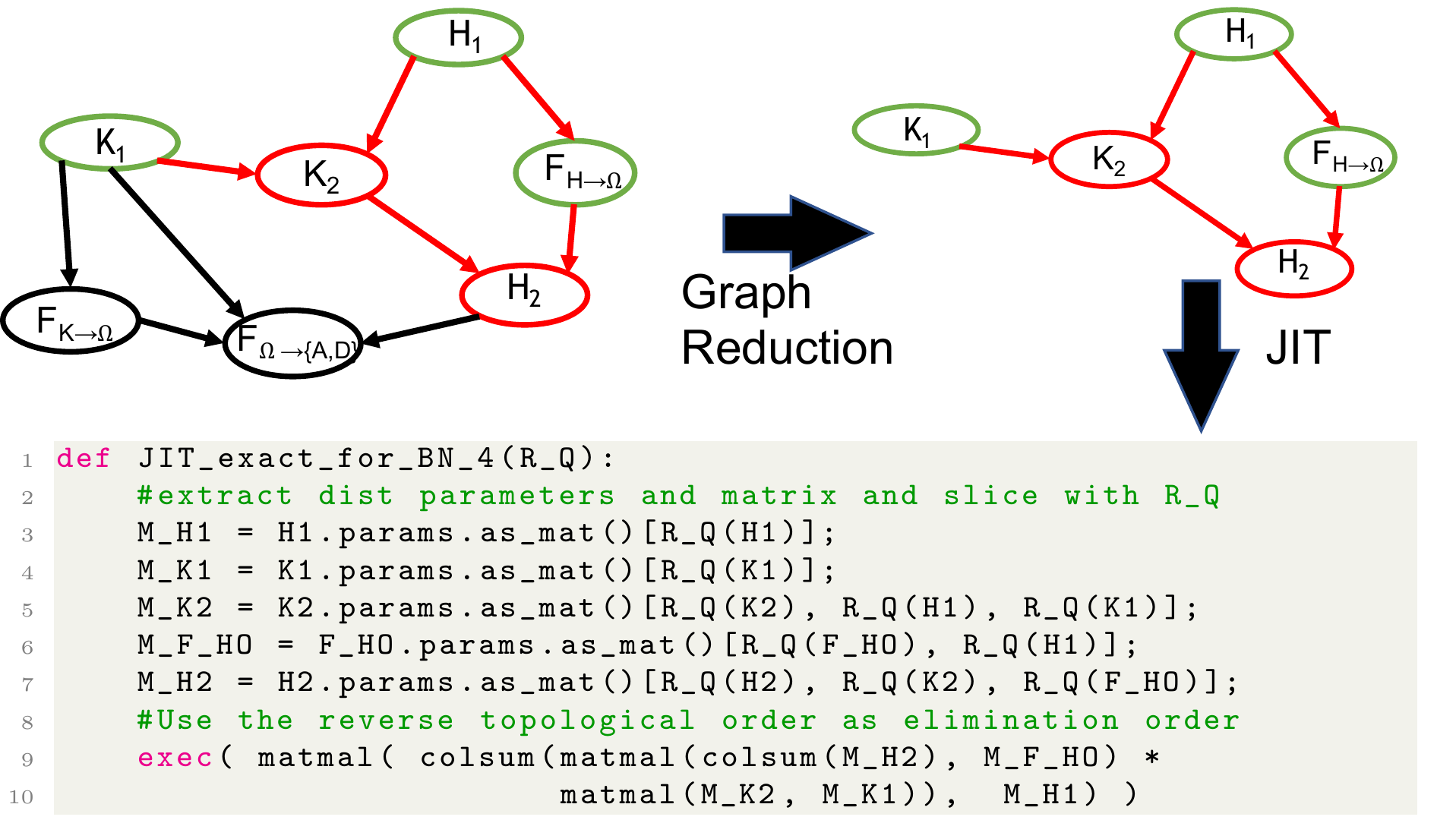}
  \vspace{-2.5em}
  \caption{ Graph reduction and the compiled program with JIT. The left image shows the graph reduction for query $K_2 \in \{10, 20\}, H_2 \in \{-1, 10\}$. The red nodes refer to the attributes in the query. All red, green nodes and the red edges form the reduced graph $G'$.}
  \vspace{-1em}
  \label{fig_RG}
\end{figure}

Progressive sampling works for general PPL programs with any distribution type. For programs restricted to categorical distributions, we can further accelerate the inference algorithm using an alternative approach: compiled variable elimination. Inspired by the impressive results of compilation for query processing~\cite{legobase_tods,dblablb,Neumann11}, we investigate the usage of just-in-time compilation (JIT) and compiler optimizations to improve inference latency. 

\smallskip
\noindent \underline{\textbf{Observation.}}
Let us revisit the example $BN_4$ built on tables $H$ and $K$, in the left image of Figure~\ref{fig_RG}. Consider a query $Q$ = ($K_1 \in \{10, 20\} \wedge H_2 \in \{-1, 10\}$), where we remove all ``black'' attributes by the graph reduction technique, based on Theorem~\ref{thm_rg}. For the ``green'' attributes, we have $R_Q(H_1) = D(H_1)$,  $R_Q(K_2) = D(K_2)$, and $R_Q(F_{H\xrightarrow{}\Omega}) = D(F_{H\xrightarrow{}\Omega})$. The variable elimination algorithm (VE) compute the probability $P_T(Q)$ based on the following equation.

\vspace{-1em}
\begin{align*}
\small
P_T(Q) = \! \! \! \! \!\! \! \! \! \!  \sum_{h_1 \in R_Q(H_1)} \! \! \! \cdots \! \! \! \! \sum_{h_2 \in R_Q(H_2) } \! \! \! \! P_T(h_1) * P_T(k_1) * \cdots * P_T(h_2|, f_{H\xrightarrow{}\Omega}, k_2)
    \label{VE} 
\end{align*}
\vspace{-1em}

This computation can be very inefficient in PPLs and repeated for estimating multiple queries. However, we observe that the VE algorithm only involves sum and product over attributes. If each variable in PPL (attribute in BN) is defined as categorical conditional distribution, they can be materialized as vectors or matrices. Thus, the VE algorithm essentially defines a program of linear algebra operations, whose execution time can be significantly enhanced by nowadays computing resource. Furthermore, we observe that the linear algebra program computing VE is fixed for a target query as long as the elimination order is fixed.

\smallskip
\noindent \underline{\textbf{JIT of VE.}}  For any query, the \textit{BayesCard} can first decide an optimal variable elimination order and then compile the learned BN from the PPL program into a static program containing only matrix or tensor operations to maximize the execution efficiency. Furthermore, this program can be re-used to infer other queries with the same reduced graph by only changing the input query regions $R_Q$ (as shown in Figure~\ref{fig_RG}). Therefore, JIT can remember the execution pattern for this query and will re-use this pattern to infer the probability of future queries for further speed-up. 

An example program showing the JIT compilation of VE on the same query $Q$ is shown in Figure~\ref{fig_RG}. Specifically, for each variable $T_i$ of PPLs in the reduced graph $G'$, the JIT program first extract the parameters of its distribution $P_T(T_i|Par(T_i))$. Since VE only supports categorical distributions, the extracted parameters of $T_i$ forms a matrix $M_{T_i}$. Next, based on the query region $R_Q$, the JIT program can further reduce $M_{T_i}$ by keeping only useful information, i.e. slicing its rows with $R_Q(T_i)$ and its columns with $R_Q(Par(T_i))$ (lines 2-6 of the code in Figure~\ref{fig_RG}). This reduction not only eliminates the redundant computation but also enables a close-form linear algebra equation.

Then, \textit{BayesCard} can determine an elimination order for these variables using the reversed topological order or standard procedure~\cite{darwiche2009modeling}. A fixed program containing only linear algebra operations can be derived, like the one in line 8, where ``\textsf{matmal}'' refers to matrix multiplication, ``\textsf{colsum}'' refers to column sum, and ``\textsf{.T}'' refers to the transpose.  At last, this generated static program can execute efficiently, thanks to the batch processing of the tensor operations with various performance tuning techniques (e.g., loop tiling, parallelization, and vectorization).
By our evaluation, such program can achieve up to two orders of magnitude speed-ups over the original VE algorithm.

\subsection{Probability inference for fanout method}
\label{sect4.3}

Previous sections discuss the process of inferring the probability $P_T(Q)$ of a query $Q$ on the table(s) $T$, represented by exactly a single BN. For a database with multiple tables, this process needs to be modified for the following two types of queries: (1) a query $Q$ on tables, that cover many BNs (i.e. $Q$ on $T = \{A,D,H,K\}$ in Figure~\ref{fig_model})-(1); (2) a query on tables, that only cover a subset of a single BN (i.e. $Q$ on $T=\{H\}$). In these cases, the \textit{BayesCard} does not contain an exact BN representing $P_T$ to estimate this query $Q$. Fortunately, based on the fanout method explained earlier in Section~\ref{sect3.2}, we can use the following theorem to calculate $P_T(Q)$, which is proposed and proved in~\cite{zhu2020flat}.

\begin{theorem}
	Given a query $Q$, let $V = \{V_{1}, V_{2}, \dots, V_{d} \}$ denote all vertices (nodes) in the join tree touched by $Q$ and let $\mathcal{V}$ denotes the full outer join of all tables in $V$. On each node $V_i$, let $F = \{ F_{A_{1}, B_{1}}, F_{A_{2}, B_{2}}, \ldots, F_{A_{n}, B_{n}}\}$, where each $(A_j, B_j)$ is a distinct join where $B_j$ is not in $Q$. Let $f = (f_1, f_2, \ldots, f_n)$ where $F_{A_{j}, B_{j}} = f_j$ for all $1 \leq i \leq n$, denote an assignment to $F$ and $\text{dlm}(f) = \prod_{j=1}^{n} \max\{f_j, 1\}$. Let
	\begin{equation}
	\small
	p_i \! = \frac{|\mathcal{V}_i|}{|\mathcal{V}|} \! \cdot \! 
	\sum\limits_{f, v}  \left( P_{\mathcal{V}_i}(Q_i  \wedge F \! = \! f \wedge F_{V_{i}, V} \! = \! v) \cdot \frac{\max\{v, 1\}}{\text{dlm}(f)} \right).
	\end{equation}
	Then, the cardinality of $Q$ is $|\mathcal{V}| \cdot \prod_{i = 1}^{d} p_i$.
\end{theorem}

In short, since all the fanout attributes involved in this computation are pre-stored in the table $V_i$ and there exists a BN for $P_{\mathcal{V}_i}$, \textit{BayesCard} can directly use this theorem for probability inference of multi-table join queries.

\smallskip
\noindent\underline{\textbf{Efficient summation computation in \textit{BayesCard}.}}
We can compute the summation $\sum_{f, v}  ( P_{\mathcal{V}_i}(Q_i  \wedge F \! = \! f \wedge F_{V_{i}, V} \! = \! v) \cdot \frac{\max\{v, 1\}}{\text{dlm}(f)} )$ over all assignments of $f$ and $v$ as efficiently as computing the probability $P_{\mathcal{V}_i}(Q_i)$ for any query. We will explain the detailed procedure for calculating $\sum_{f \in D(F)} P_T(Q, F=f) * f$ using progressive sample and complied variable elimination, where $D(F)$ denotes the domain of unique values in $F$. Then, this procedure can naturally generalize to more complex cases. 

Our calculation procedure is motivated by the Bayesian rule, that $P_T(Q, F=f) = P_T(F=f|Q) * P_T(Q)$. We observe that $P_T(Q)$ is a fixed value independent of $F$ because the fanout attributes are artificial attributes that will not be involved in $Q$. Furthermore, by property of BN, we know that $P_T(F|Q) = P_T(f|R_Q(Par(F)))$, so can derive the following equation. It spots a common term $P_T(Q)$ so the calculation can avoid repeatedly computing $P_T(Q)$.
\begin{equation*}
\sum_{f \in D(F)} P_T(Q, F=f) * f = P_T(Q) * \left(\sum_{f \in D(F)} P_T(f|R_Q(Par(F)))*f \right)
\end{equation*}

\noindent \textbf{Progressive sampling.} Recall in Section~\ref{sect4.2}, \textit{BayesCard} estimates $P_i = P_T(T_i|R_Q(Par(T_i)))$ by making progressive samples of $R_Q$ and approximate the $P_T(Q)$ as $\prod P_i$. After finishing estimating $P_T(Q)$ with sample $S$, \textit{BayesCard} can directly estimate $\sum_{f \in D(F)} \break P_T(f|R_Q(Par(F)))*f$ using the same sample $S$, i.e. as $\sum_{f \in S[F]} \break \hat{P}_T(f| S[Par(F)])*f$. The final result can be achieved by multiplying these two terms together.

\noindent \textbf{Compiled variable elimination.} Recall in Section~\ref{sect4.2}, \textit{BayesCard} can specify a particular elimination order by choosing the fanout variable $F$ as the last variable to eliminate. Using PPL, the intermediate result after each elimination step is materialized as a distribution. Therefore, before the last elimination step of VE algorithm for computing $P_T(Q)$, \textit{BayesCard} can store the intermediate result, which represents the conditional distribution $P_T(F|Q)$. Then, the summation $\sum_{f \in D(F)} P_T(f|R_Q(Par(F)))*f$ equals to $P_T(F|Q) \cdot D(F)$, where $\cdot$ denotes the vector dot product. Therefore, similar to computing $P_T(Q)$, this process only involves linear algebra operations, which can be compiled and efficiently calculated using JIT.

\section{Model construction of \textit{BayesCard}}
\label{sect5}
In this section, we explain how \textit{BayesCard} constructs an ensemble of BNs for a multi-table database. Specifically, Section~5.1 first introduces the BN ensemble construction method with budget, which clusters all tables in the database into several groups and builds a single BN on each group of tables. Then, Section~5.2 introduces some optimizations for building a single BN using PPLs. Finally, Section~5.3 shows how to incrementally update the BN model.

\begin{figure}[t]
  \centering
  \vspace{-1.5em}
  \includegraphics[width=8.5cm]{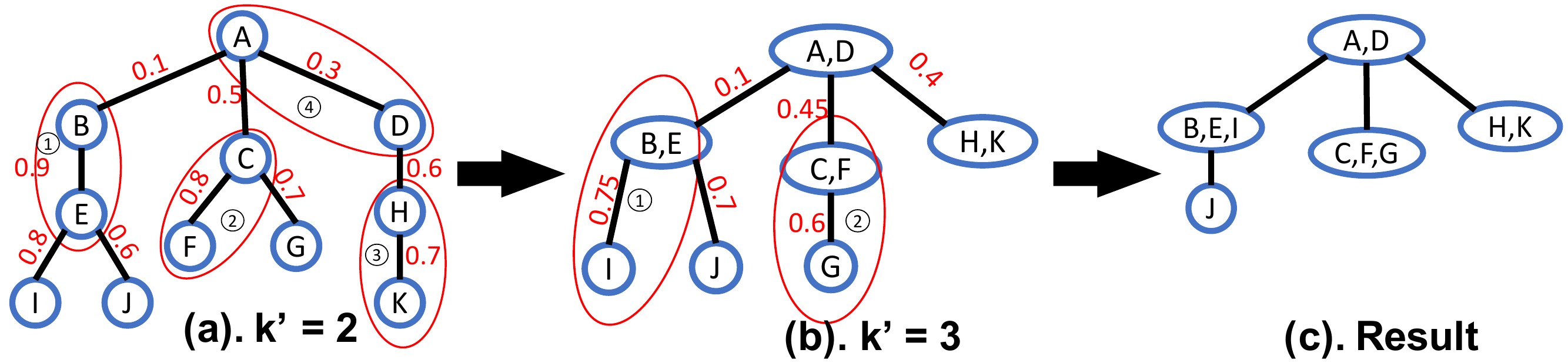}
  \vspace{-1.5em}
  \caption{\textit{BayesCard} ensemble learning algorithm demo.}
  \vspace{-2em}
  \label{PRM_learn}
\end{figure}

\subsection{Ensemble construction with budget}
\label{sect5.1}


\noindent\underline{\textbf{Main idea.}}
Consider the example database in Figure~\ref{PRM_learn} with 11 tables $A, B,  \ldots, K$  forming a join tree, where each node represents a table and each edge represents a possible join between two tables. A previous approach~\cite{deepDB} suggests to create every possible two-table join results, examine the level of dependence between attributes across the two, and determine whether to create one large model on their full outer join table or two separate models. Since generating the full outer join of multiple tables could require exponential memory, this approach normally can not explore the possibility of creating a model on the join of more than three tables. 

Another approach~\cite{NeuroCard} generates an unbiased sample $S$ on the full outer join of all tables in the schema and builds a single large model on $S$ directly. As the resulting model is built on all attributes in the database, the model construction and the probability inference can be very inefficient. Moreover, the size of $S$ is relatively small with respect to the full outer join size, suggesting a large amount of information loss, so the learned model on $S$ might not accurately represent the actual data distribution.

In order to balance the estimation accuracy and inference efficiency, we want to explore the full possibility of learning different BN ensembles such that the number of joined tables in each BN is no more than a threshold. Therefore, the resulting ensemble should capture as much dependence between tables as possible and simultaneously keep each BN in this ensemble as small as possible. 



\begin{algorithm}[t]
\small
\caption{BN Ensemble Construction Algorithm}
\label{PRM_learn_algo}
\begin{flushleft}
\textbf{Input}: a DB schema with n tables $T_1, \cdots, T_n$ and a budget $k$
\end{flushleft}
\begin{algorithmic}[1]
\State Create the join tree $\mathcal{T} = (V, E)$ for the schema
\State Generate unbiased samples $S$ for full outer join of the entire schema
\State Initialize a dependence matrix $M \in \mathbb{R}^{n \times n}$
\For{Each pair of tables $e = (T_i, T_j)$}
    \State Calculate the RDC dependence level scores between all attributes in $T_i$ and attributes in $T_j$
    \State $w_e$ $\gets$ average RDC scores
    \EndFor
\If{$k = 1$} \State \textbf{return} $\mathcal{T}$ and learn a single PRM for each table
\EndIf
\For{$k' \gets 2, \cdots, k$}
    \State Sort $E$ in decreasing order based on $w_e$.
    \For{$e = (u, v) \in E$}
        \If{$u$ and $v$ contain exactly $k'$ tables in total} 
        \State Update $T$ by contracting nodes $u, v$ to a single node $\{u, v\}$
        \EndIf
    \EndFor
\EndFor
\State \textbf{return} $\mathcal{T}$ and learn a single PRM for each node in $\mathcal{T}$
\end{algorithmic}
\end{algorithm}

\smallskip
\noindent\underline{\textbf{Algorithm description.}}
The details of the ensemble construction algorithm is given in Algorithm~\ref{PRM_learn_algo}.
First, we define the budget $k$ such that a single BN model can only be constructed on (a sample of) the full outer join of no more than $k$ tables. The budget $k$ is a hyper-parameter decided by the dataset, system, and computing resource. The algorithm generally works as follows:

1) \textit{Computing dependency between tables (lines 1-7).} Given a tree-structured join schema $\mathcal{T}$, we first generate the unbiased sample $S$ of the full outer join of all tables according to~\cite{Sampling}. Specifically, the join tree is regarded as a rooted tree and samples $S$ are obtained by scanning all tables in $\mathcal{T}$ in a bottom-up manner. Then, we calculate the randomized dependence coefficient, i.e., RDC value~\cite{rdc}, between each pair of join tables using $S$. The detailed computation method is given in Appendix~B of our technical report~\cite{wu2020bayescard}. In Figure~\ref{PRM_learn}, the RDC value is shown as red numbers on each edge.

2) \textit{Contracting nodes (lines 8-18).} Intuitively, we would like to build a model on the full outer join of tables with high dependency. We can iteratively contract the nodes (tables) with high RDC value in $\mathcal{T}$ in a greedy manner. Let $k' = 2$ at the beginning. In each iteration, if $k' \leq k$, we first sort all edges $e = (u,v)$ (joins) in a descending order based on their RDC values. According to this edge order, we aggregate $u, v$, i.e. two endpoints of edge $e$, into a single node if they contain exactly $k'$ tables in total and update the RDC values of $e$ accordingly, whose details is given in Appendix~B of our technical report~\cite{wu2020bayescard}. We iterate this process until $k' = k$, and in the end, we obtain a tree where each node contains at most $k$ tables. For example, in Figure~\ref{PRM_learn}, let the budget $k = 3$. In the first iteration where $k' = 2$, the algorithm considers joining two tables together. The edge $(B, E)$ has the highest RDC value, so $B$ and $E$ are aggregated in the first step (\textcircled{1} in Figure~\ref{PRM_learn}(a)). After the first iteration, the join schema $\mathcal{T}$ has been transformed into a new tree in Figure~\ref{PRM_learn}(b). Similarly, in the second iteration where $k' = 3$, the node $\{B, E\}$ is first merged with the node $I$. Finally, the join tree is transformed to a tree in Figure~\ref{PRM_learn}(c).

3) \textit{Building BNs (line 19).} In the end, \textit{BayesCard} will construct a single BN model on (a sample of) the full outer join of tables within each node and fanout attributes will be added accordingly.

\smallskip
\noindent\underline{\textbf{Time Complexity analysis.}}
As shown in~\cite{Sampling}, creating the samples $S$ on the full outer join of tables $T_1, \cdots, T_n$ takes $O(\sum_{i=1}^{n}|T_i|)$ time.
Let $m$ be the attribute number in the full outer join of the tables. Calculating the pairwise RDC values takes $O(m^2|S|\log |S|) $. The rest of Algorithm~\ref{PRM_learn_algo} takes $O(kn^2)$ time since the algorithm terminates in $k$ iterations and in each iteration we only need to check the tables defined by two endpoints of each edge, which is at most $n^2$. Thus, the whole time complexity is $O(\sum_{i=1}^{n}|T_i| + m^2|S|\log |S| + kn^2)$.

\subsection{Single model construction optimizations}
\label{sect5.2}
The structure learning process, i.e., learning the causal structure from data of a single BN, is an NP-hard combinatorial optimization problem~\cite{34}. Current structure learning algorithms supported by PPLs either produce a general DAG structure or a simplified tree structure. We show optimization techniques for them as follows:

\smallskip
\noindent \underline{\textbf{Optimization for DAG structure learning algorithms.}}
The exact DAG structure learning algorithms explore the super-exponential searching space of all possible DAGs and select the best candidate~\cite{MDL, BIC, BDeu, A-start}). The learned structure is accurate but inefficient, which only scales to tens of attributes. Approximate methods limit the searching space with local heuristic (i.e. \emph{greedy} algorithms~\cite{greedysearch, 36, 37}), but they may produce inaccurate results.
Based on PPLs, \textit{BayesCard} supports pre-specifying sub-structures before running the \emph{exact} and \emph{greedy} structure learning algorithms, which limits the DAG searching space and makes the structure learning much more efficient. Specifically, practical databases generally exist attributes with \emph{functional dependencies}~\cite{fan2010discovering} or obvious causal relations between attributes, such as one's ``age'' determining one's ``school level''. First, users of \textit{BayesCard} can use their ``expert knowledge'' to pre-specify certain causal structures for subsets of attributes. Then, the PPLs within \textit{BayesCard} can define the variables corresponding to these attributes, and condition the variables with each other according to the pre-specified structure. At last, \textit{BayesCard} can rely on the existing algorithms to construct the remaining causal structure on these variables. Since the algorithms are forced to maintain these sub-structures, the number of qualified DAG candidates is significantly curtailed, making the structure learning process more efficient without loss in accuracy.

\smallskip
\noindent \underline{\textbf{Optimization for tree structure learning algorithms.}}
The tree structure learning algorithm learns a tree structure such as \emph{Chow-Liu tree}~\cite{23}, which sacrifices accuracy for efficiency.
\textit{BayesCard} can also improve the accuracy of a learned structure using the aforementioned ``expert knowledge'' after running the \emph{Chow-Liu tree} algorithm. This efficient algorithm forces the learned BN structure to be a tree, which could contain ``false'' causality or miss important attribute dependence. For example, intuitively we know that the number of ``children'' raised by someone is largely dependent on one's ``income'' and one's ``marital status'', which can not be captured simultaneously by the tree BN, since one node is only allowed to have one parent. Thus, after the structure is learned, \textit{BayesCard} can add the edge from ``Income'' to ``Children'' to improve its accuracy. With PPLs, only the parameters of the affected sub-structure (the ``Children'' variable in this example) need to be updated.

\subsection{Model updates}

Most of the practical databases update their data frequently, requiring the cardinality estimators to adjust their underlying models dynamically~\cite{wang2020ready}. When the data distribution changes, \textit{BayesCard} can update its underlying BNs very efficiently. Specifically, the learned structure of BN captures the \emph{intrinsic} causal pattern of the attributes, which is not likely to change even in the case of massive data updates. Therefore, in most cases, \textit{BayesCard} can preserve the original BN structure and only \emph{incrementally} update its distribution parameters. Such parameter updates are extremely efficient using MLE in PPLs. By our testing, it generally takes less than one second for an insertion or deletion of a thousand tuples. In some rare cases involving the insertion or deletion of attributes, a new BN structure should be constructed. Even in this case, the causal pattern of the original attributes is largely preserved. Therefore, \textit{BayesCard} can pre-specify some sub-structures and learn the new structure efficiently using the methods stated in the previous section.

\section{Analysis of BayesCard}
In this section, we analyze and demonstrate that \textit{BayesCard} satisfies the ADS criteria from all aspects, as shown in Table~\ref{ADSsummary}.

\noindent\textbf{Algorithm.} A BN with exact learned structure can losslessly capture the data distribution, a.k.a. near-perfect \emph{estimation accuracy} for all queries. We show empirically that even with an approximate tree structure, \textit{BayesCard} can achieve comparable or better accuracy than the current SOTA methods. The \emph{inference latency} of \textit{BayesCard} is roughly 1ms per query (close to Histogram method), thanks to our novel inference algorithms. Furthermore, as explained in Section~\ref{sect4}, \textit{BayesCard} can learn a compact structure of small \emph{model size} with fast \emph{training and update time}.

\noindent\textbf{Data.} Every dataset contains an inherent causal pattern, which can be discovered by \textit{BayesCard}. Building upon this structure, \textit{BayesCard} can represent its PDF accurately and efficiently. Specifically, the variables in PPL can characterize most data \emph{distribution} types with varied \emph{domain size}. \emph{Attribute correlation} is merely a manifestation of the underlying causal pattern, which can be accurately represented. Moreover, for data with more attributes (larger \emph{scale}), the proposed \emph{graph reduction} inference technique can reduce a larger graph into a much smaller one for efficient inference. Therefore, the inference latency is also stable for various data settings.

\noindent\textbf{System.} Both the structure and the distribution parameters of \textit{BayesCard} model are \emph{interpretable} and \emph{debuggable}. Specifically, a DB expert can verify a learned structure based on his prior knowledge of data causality (functional dependency in DBs), and validate the learned parameter using basic probability rules (non-negative and sum to one). Since the probability inference of \textit{BayesCard} follows the Bayesian rule, its performance is logical and \emph{predictable}. Furthermore, the compiled VE does not contain any stochasticity, so the users' error is \emph{reproducible}.

\section{Experimental Results}
\label{sect6}
In this section, we empirically demonstrate the superiority of our \textit{BayesCard} over other \CE methods. In the following,
Section~\ref{sect6.1} first introduces the experimental setups.
Next, Section~\ref{sect6.2} thoroughly compares different \CE methods in terms of the \emph{ADS} criteria on single table datasets.
Then, Section~\ref{sect6.3} evaluates the performance and end-to-end query plan execution time on multi-table datasets.
At last, Section~\ref{sect6.4} performs ablation studies on our proposed algorithms and optimizations in \textit{BayesCard} method.

\subsection{Experimental setups}
\label{sect6.1}

\underline{\textbf{\CE methods to compare with.}}
We compare our \textit{BayesCard} framework with the following \CE methods, including both traditional methods widely used in DBMS and four existing SOTA DL-based methods. For each ML-based \CE method, we adopt the authors’ source code and apply the same hyper-parameters as used in the original paper.

\textit{1). Histogram} is the simplest \CE method widely used in DBMS such as Postgres~\cite{postgresql}.

\textit{2). Sampling} has been used in DBMS such as MySQL~\cite{mysql}. In our testing, we randomly sample $1\%$ of all tuples for \textsf{CardEst}.

\textit{3). Naru/NeuroCard}~\cite{naru,NeuroCard} are \textsf{DAR}-based \CE methods for single table and multi-table join queries, respectively. 

\textit{4). DeepDB}~\cite{deepDB} is a SPN-based \CE method. 

\textit{5). FLAT}~\cite{zhu2020flat} is an FSPN-based \CE method. 

\textit{6). MSCN}~\cite{MSCN} is the SOTA query-driven \CE method. For each dataset, we train it with $10^5$ queries generated in the same way as the workload.

Our \textit{BayesCard} framework subsumes BNs with various combination of structure learning and inference algorithms as described in previous sections.  In Section~\ref{sect6.2} and~\ref{sect6.3}, we use an exemplary BN with \emph{Chow-Liu} tree structure learning algorithm and \emph{compiled variable elimination} inference algorithm with graph reduction optimizations. The comparison of different BNs realizable in \textit{BayesCard} and controlled ablation studies are deferred to Section~\ref{sect6.4}.





\smallskip
\noindent\underline{\textbf{Datasets and query workloads.}}
Our single table experiments are performed on three datasets: 

\noindent 1).\textbf{DMV} dataset is a real-world dataset consisting of 11,575,483 tuples of vehicle registration information in New York. We use the same attributes as in~\cite{naru, wang2020ready}. 

\noindent 2). \textbf{CENSUS} dataset contains population survey by
U.S. Census Bureau conducted in 1990. This dataset has 2,458,285 tuples and 68 attributes, containing highly correlated attributes. Based on RDC test~\cite{rdc}, we find that more half of the attributes are highly correlated with at least one other attribute. This dataset is very large in scale and has very complicated distribution.

\noindent 3) \textbf{SYNTHETIC} datasets are a collection of human-generated datasets with varied data distribution skewness, attributes correlation, domain size and number of attributes. We generated these datasets using the similar approach as a recent benchmark study~\cite{wang2020ready}. They are used to evaluate models' stability w.r.t.~changes in data.

For each dataset, we generate $1,500$ selection queries as workload. For each query $Q$, first we select a subset of attributes as filter attributes of $Q$. For each selected attribute $c$, if it represents a continuous variable, we uniformly generate two values ($v_1, v_2$) from its value domain and then add the filter predicate ``$v_1 \leq c \leq v_2$'' to $Q$. Otherwise, if $c$ is a categorical variable, we uniformly sample $k$ unique values\{$v_1, v_2, \cdots, v_k$\} from its domain and place a predicate ``$c$ \textsc{ IN } \{$v_1,\cdots, v_k$\}'' in $Q$.

\noindent 4). \textbf{Multi-table IMDB:}
We conduct the multi-table experiment on international movie database (IMDB) benchmark. Prior work~\cite{howgoodare} claims that this DB contains complicated data structure and establishes it to be a good test benchmark for cardinality estimators. We use \emph{JOB-light} benchmark query workload with 70 queries proposed in the original paper~\cite{howgoodare} and create another workload of 1500 \emph{JOB-comp} with more \underline{comp}rehensive and \underline{comp}licated queries.

\textit{JOB-light}'s IMDB schema contains six tables (\textsl{title}, \textsl{cast\_info}, \textsl{movie\_info}, \textsl{movie\_companies}, \textsl{movie\_keyword}, \textsl{movie\_info\_idx}) and five join operations in total where every other tables can only join with the primary table ``title''. Each \textit{JOB-light} query involves 3-6 tables with 1-4 filter predicates. The filter variety is not very diverse with equality filters on all attributes but the ``title.production\_year'' attribute only. In addition, \textit{JOB-light}'s workload only contains 70 queries, which is not enough to account for the variance in model prediction. Thus, we synthesize 1,500 \emph{JOB-comp} queries based on the schema of \emph{JOB-light} with more number of filter predicates per query. Each \emph{JOB-comp} query involves 4-6 tables with 2-7 filter predicates. The queries are uniformly distributed to each join of 4-6 tables. After determining the join graph, the filter predicates selection process is similar as in single table cases.

\smallskip
\noindent\underline{\textbf{Evaluation metric:}} We use the Q-error as our evaluation metrics, which is define as follow:
\begin{equation*}
    \textbf{Q-error} = max(\frac{\text{Estimated Cardinality}}{\text{True Cardinality}}, \frac{\text{True Cardinality}}{\text{Estimated Cardinality}})
\end{equation*}
This evaluation metric is well recognized in DBMS community and widely used in recent papers on cardinality estimation~\cite{deepDB, naru, NeuroCard, 2001SigmodGreedy, tzoumas2011lightweight}. We report the \textbf{50\%}(median), \textbf{90\%}, \textbf{95\%} and \textbf{100\%}(worst) Q-error quantiles as evaluation of estimation accuracy. 

\noindent\underline{\textbf{Experimental environment:}}
All models are evaluated on Intel(R) Xeon(R) Platinum 8163 CPU with 64 cores, 128GB DDR4 main memory, and 1TB SSD. For a fair comparison, we compare the model inference latency on CPU only since apart from the DAR model (\textit{Naru} and \textit{NeuroCard}) and \textit{MSCN}, the rest methods' inference algorithms do not support GPU.

\begin{table}[t]
  \caption{Performance of \CE algorithms on single tables.}
  \vspace{-1em}
	\resizebox{\columnwidth}{!}{
		\begin{tabular}{c|c|cccc|c}
			\hline
			Dataset& Method & $50\%$ & $90\%$ & $95\%$ & $100\%$ & Latency (ms) \\ \hline
			\multirow{7}{*}{DMV} 
			&\textbf{\textit{BayesCard}} &\firstcell{1.001} &\firstcell{1.024} &\secondcell{1.049} &\secondcell{7.641} &\thirdcell{2.1} \\ \cline{2-7}
          &Histogram &1.318 & 12.32 & 143.6 & $1\cdot 10^4$ &\firstcell{0.1} \\ \cline{2-7}
          &Sampling & 1.004 & 1.052 & 1.140 & 143.0 & 79 \\ \cline{2-7}
          &Naru & \thirdcell{1.003} & \secondcell{1.026} & \firstcell{1.035} & \firstcell{5.500} & 86 \\ \cline{2-7}
          &DeepDB & 1.006 & 1.124 & 1.193 & 108.1 & 5.1 \\ \cline{2-7}
          &FLAT & \firstcell{1.001} & \thirdcell{1.028} & \thirdcell{1.066} & \thirdcell{11.37} & \secondcell{0.6} \\ \cline{2-7}
          &MSCN & 1.210 & 2.263 & 4.507 & 151.8 & 3.4 \\
			\thickhline
			\multirow{7}{*}{CENSUS} 
			&\textbf{\textit{BayesCard}} & \firstcell{\textbf{1.063}} & \firstcell{\textbf{1.484}} & \firstcell{\textbf{2.052}} & \firstcell{\textbf{227.5}} & \secondcell{2.4} \\  \cline{2-7}
			&Histogram &5.561 &259.8 &$5\cdot 10^4$ & $5\cdot 10^5$ & \firstcell{\textbf{0.2}}\\ \cline{2-7}
          &Sampling & \secondcell{1.130} & \secondcell{1.412} & 374.2 & \thirdcell{1703} & 113 \\ \cline{2-7}
          &Naru &\thirdcell{1.229} & \thirdcell{2.210} & \secondcell{7.156} & \secondcell{1095} & 129 \\\cline{2-7}
          &DeepDB & 1.469 & 6.295 & 178.21 & $1\cdot 10^4$ & 25 \\ \cline{2-7}
          &FLAT & 1.452 & 6.326 & \thirdcell{174.93} & $1\cdot 10^4$ & 25 \\ \cline{2-7}
          &MSCN & 2.700 & 15.83  &$1\cdot 10^4$ & $1\cdot 10^5$ & \thirdcell{4.8} \\
			\hline
	\end{tabular}}
	\vspace{-0.5em}
	\label{tab: exp-single}
\end{table}

\subsection{Model evaluation on single tables}
\label{sect6.2}
In this section, we compare the performance of \CE methods in terms of \emph{Algorithm} and \emph{Data} criteria.


\smallskip
\noindent\underline{\textbf{Algorithm criteria.}}
We evaluate the \CE methods from four aspects: estimation accuracy, inference latency, model size and training time, and updating effects.

\noindent\textbf{Estimation accuracy:}
The estimation accuracy on two real-world single table datasets is reported in Table~\ref{tab: exp-single}, where the color shade in each cell corresponds to the rank among different \CE methods. When compared with traditional models (\textit{Histogram} and \textit{Sampling}), \textit{BayesCard} achieves $1$--$3$ order of magnitude higher accuracy than both models. When compared with DL-based methods (\textit{Naru}, \textit{DeepDB} and \textit{FLAT}), \textit{BayesCard} has comparable or better estimate accuracy on DMV dataset, but significantly more accurate on CENSUS dataset. This is because these DL models can accurately represent the data distribution of DMV, which contains relatively less attribute correlation and fewer number of attributes. CENSUS, however, contains seven times larger number of attributes with more complex attribute correlations.
As the learning space grows exponentially with the number of attributes, \textit{Naru}'s accuracy dropped significantly. 
For \textit{DeepDB} and \textit{FLAT}, their SPN or FSPN structure can not well capture the data distribution in presence of a large number of highly correlated attributes, so their performance also heavily degrades. 

\noindent\textbf{Inference latency:} 
As shown in Table~\ref{tab: exp-single}, apart from \textit{Histogram}, which leverages the attribute independence assumption for fast inference, \textit{BayesCard} generally attains the best ($1$--$2$ orders of magnitude) inference latency among the result methods. 
Worth noticing that we observe significant increase in latency from DMV to CENSUS datasets for all methods except for \textit{BayesCard}. \textit{BayesCard}'s inference time appears to be insensitive to the number of attributes, mainly because the novel \emph{graph reduction} technique can reduce a large CENSUS attribute graph to a much smaller one for inference.

\begin{figure}[t]
  \centering
  \includegraphics[width=\linewidth]{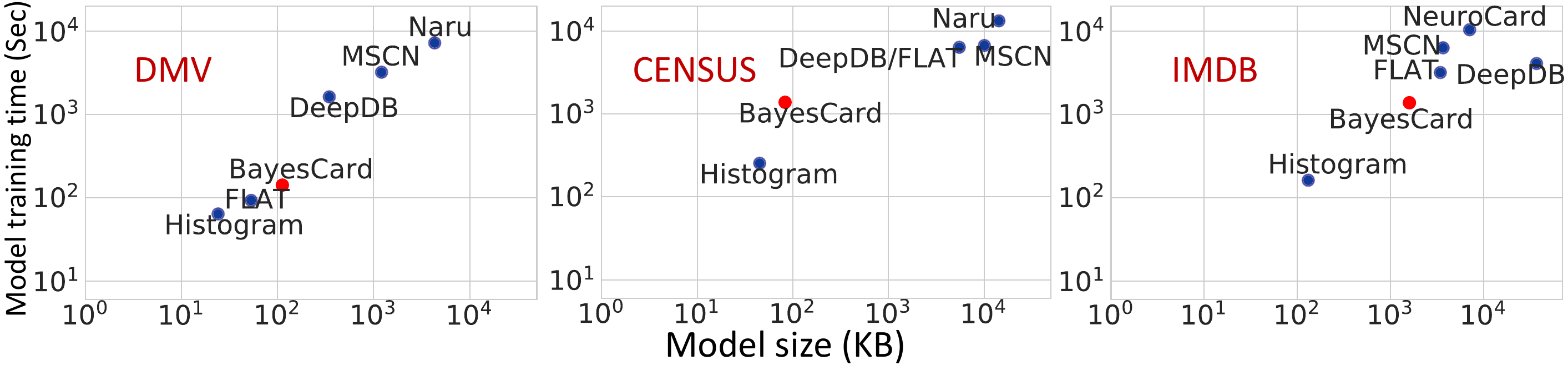}
  \vspace{-2.5em}
  \caption{ Model storage and training time. }
  \vspace{-1em}
  \label{model_size}
\end{figure}

\noindent\textbf{Model size and training time:} As shown in Figure~\ref{model_size}, apart from the traditional methods, \textit{BayesCard} achieves the smallest model size with the fastest training time because the causal pattern of datasets enables a compact representation of data distribution. Worth noticing that Sampling is a model-free method that does not have model size or training time, so we do not include it in the figure.

\noindent\textbf{Updating time:} 
We evaluate each method's updating effects by following a similar experimental setup of prior work~\cite{wang2020ready}. 
Specifically, we create a copy of the original DMV dataset and sort the tuples based on the value of each column in an ascending order. Then, we take the first $20\%$ of the data to train a stale model and use the rest $80\%$ as data insertion updates. This procedure will make sure that the training dataset has different data distribution than the testing dataset; otherwise, the stale model would perform well without model updates. Then, after the model finishes the updating process, we test the model using the original query workload same as in Table~\ref{tab: exp-single} and report their $95\%$ q-errors and total update time in Figure~\ref{update}. Here, we refrain from comparing with the query-driven method \textit{MSCN} because it requires a new query workload to update its model, which is unavailable in our experimental settings.

\begin{table}[t]
  \caption{Performance of model updates of different \CE methods on DMV. The baseline q-error is the 95\% q-error quoted from Table~\ref{tab: exp-single} for comparison.}
  \vspace{-1em}
	\resizebox{0.9\columnwidth}{!}{
		\begin{tabular}{c|c|c|c|c|c}
			\hline
			Method & \textbf{\textit{BayesCard}} & Histogram & Naru & DeepDB & FLAT \\ \hline
			baseline q-error & 1.049 & 143.6 &1.035 & 1.193 &1.066\\ \hline
			95\% q-error & \textbf{1.049} & 143.6 & 14.79 & 18.83 & 1.451\\ \hline
			Update time (s) &103 &\textbf{25} & 1980 & 142 & 257 \\ 
			\hline
	\end{tabular}}
	\vspace{-1.5em}
	\label{update}
\end{table}

\textit{BayesCard}, \textit{Histogram}, and \textit{DeepDB} all preserve the original structure and only incrementally update the parameters, so in general, they have the fastest update time. Among them, \textit{Histogram} has the least amount of parameters to update, so it has the best update time. We use the method described in the original paper~\cite{zhu2020flat} to update \textit{FLAT}, which generates new sub-structures to fit the inserted data distribution, so it is slightly slower than the previous three. \textit{Naru} uses the incoming data to fine-tune its pre-trained DNNs for three epochs, which is significantly slower than others. 

After the model updates, we observe that \textit{BayesCard} has no drop in estimation accuracy, whereas the deep probabilistic models have degraded performance. The reasons can be summarized as follow: (1) \textit{BayesCard}'s structure captures the data causal pattern, which often does not change after update; (2) \textit{DeepDB}'s preserved structure is not robust against data distribution changes; (3) fine-tuning the \textit{Naru}'s underlying DAR model overfits the information from the $20\%$ previously trained data, leading to degraded performance.

\noindent\textbf{\textit{Summary:}}
\textit{\textit{BayesCard} attains comparable or better estimation accuracy, lower inference latency, smaller model size, less training and update time than DL-based models. In addition, \textit{BayesCard} is 1-3 orders of magnitude more accurate than traditional methods.}

\smallskip
\noindent\underline{\textbf{Data criteria.}} 
We evaluate the stability of \CE methods in terms of \emph{Data} criteria from four aspects: data distribution, attribute correlation, domain size, and number of attributes.

SYNTHETIC datasets are generated using the similar approach in a recent benchmark study~\cite{wang2020ready}. Specifically, suppose we would like to generate a table $T$ with attributes $\{T_1,\ldots,T_n\}$ and $10^6$ tuples, where is the $n$ denotes the number of attributes (\emph{scale}). We generate the first column for $T_1$ using a Pareto distribution (using scipy.stats.pareto function), with a controlled skewness $s$ and domain size $d$. For each of the rest attribute $T_i$, we generate a column based on a previous attribute $T_j$ with $j<i$, to control the correlation $c$. For each tuple ($t_1, \ldots, t_n$) in $T$, we set $t_i$ to $t_j$ with a probability of $c$, and set $t_i$ to a random value drawn from the Pareto distribution with the probability of $1-c$. 

\begin{figure}[!t]
  \centering
  \includegraphics[width=\linewidth]{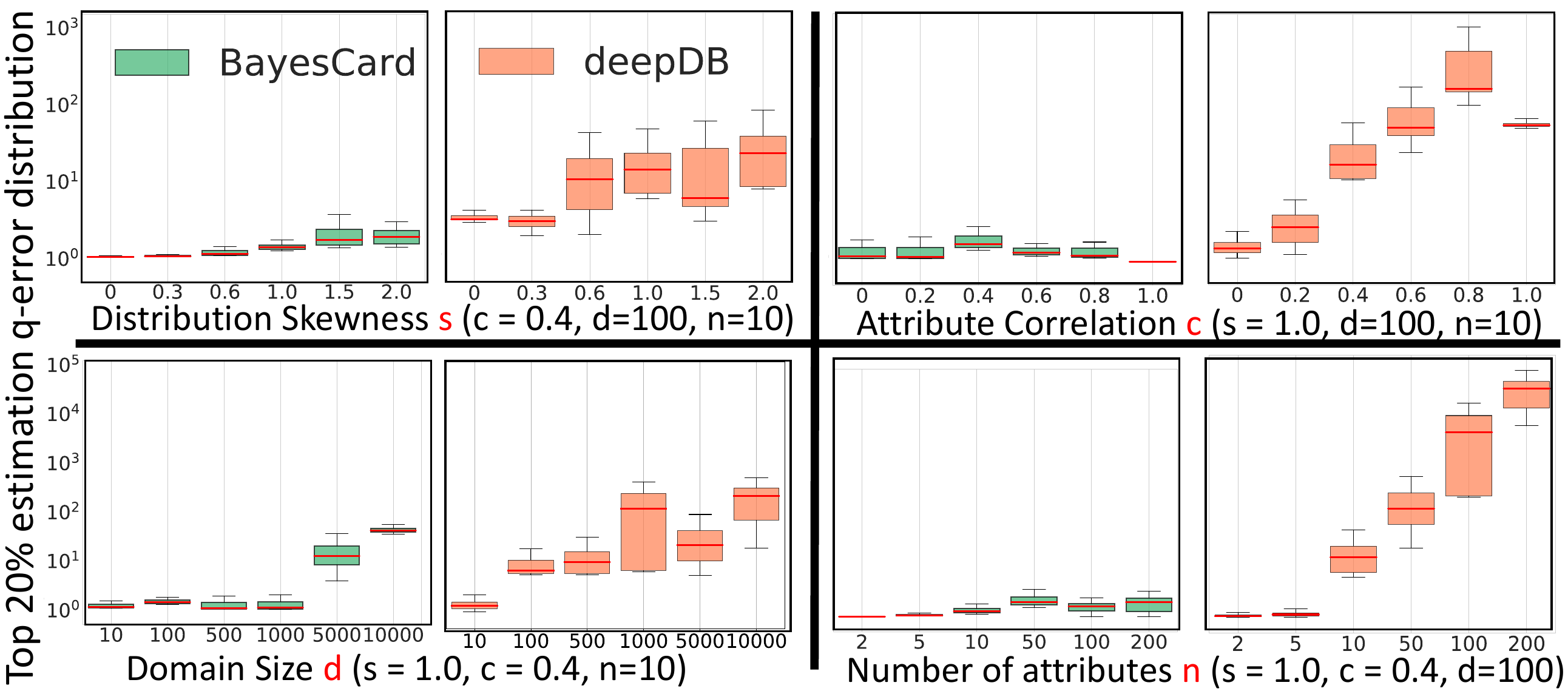}
  \vspace{-2em}
  \caption{ Comparing \textit{BayesCard} and DeepDB's stability. }
  \vspace{-2em}
  \label{synth}
\end{figure}

The experimental results on SYNTHETIC are shown in Figure~\ref{synth}. Due to space limit, we only plot the comparison results between \textit{BayesCard} and \textit{DeepDB} on the estimation accuracy metric. The additional experimental results are reported in the appendix of the technical report~\cite{wu2020bayescard}. We summarize our observations as follows.

\noindent\textbf{\text{Distribution (s):}} Similar to the previous study~\cite{wang2020ready}, we find that increasing the Pareto distribution skewness severely degrades the performance of \textit{Naru} and \textit{Sampling} methods, but has only mild effect on \textit{BayesCard} and other methods. This is because \textit{BayesCard}, \textit{Histogram}, \textit{FLAT}, and \textit{DeepDB} all use (multi-)histograms to represent distributions, which are robust against distribution changes. 

\noindent\textbf{\text{Correlation (c):}} The increase in $c$ has no impact on \textit{BayesCard}, mild impact on \textit{Sampling}, \textit{Naru}, \textit{FLAT} and \textit{MSCN}, and severe impact on \textit{Histogram} and \textit{DeepDB}, which make local or global attribute independence assumptions. \textit{BayesCard} is able to capture the causal pattern of the datasets, and thus can represent any attribute correlation accurately.

\noindent\textbf{\text{Domain (d):}} The increase in the domain size degrades the estimation accuracy for all methods, because increasing $d$ may increase the data complexity exponentially as there are $d^n$ possible values that a tuple can take. Fortunately, except for \textit{Naru}, the degrades in accuracy are within a reasonable range for all other methods. 

\noindent\textbf{\text{Scale (n):}} Similar to domain size, increasing the number of attributes also increases the data complexity exponentially, and thus we expect to see a decrease in accuracy for all methods. Surprisingly, the performance of \textit{BayesCard} was not affected by $n$ at all. This is owing to the graph reduction technique, which significantly reduces the number of attributes involved during inference. This technique not only improves the inference latency but also increases the estimation accuracy as potential modeling errors on the reduced attributes are also eliminated.

Apart from estimation accuracy, \textit{BayesCard} also maintains very stable and robust performance in terms of inference latency, model size, and training time, which is analyzed in Appendix~C~\cite{wu2020bayescard}.

\noindent\textbf{\textit{Summary:}}
\textit{\textit{BayesCard} is much more stable and robust than other \CE methods for datasets with various settings of data.}

\subsection{Model performance on multi-table dataset}
\label{sect6.3}

As reported in Table~\ref{tab: exp-multi} and Figure~\ref{model_size}, \textit{BayesCard} achieves comparable performance with the current SOTAs on the two query workloads of the IMDB dataset and preserves its superior inference latency, lightweight model storage, and fast training. Specifically, the estimation accuracy of \textit{BayesCard} is comparable to \textit{NeuroCard}, slightly better than \textit{DeepDB}, and slightly worse than \textit{FLAT}, but with up to $60\times$ smaller model size, and $10\times$ faster training and inference.

\begin{table}[t]
 	\vspace{-0.5em}
 	\caption{Performance of cardinality estimation algorithms on IMDB datasets with two query workloads.}
 	\vspace{-1em}
	\resizebox{\columnwidth}{!}{
		\begin{tabular}{c|c|cccc|c}
			\hline
			Workload& Method & $50\%$ & $90\%$ & $95\%$ & $100\%$ & Latency (ms) \\ \hline
			\multirow{7}{*}{JOB-light}
          &\textbf{\textit{BayesCard}} & \secondcell{1.300} & \thirdcell{3.534} & \thirdcell{4.836} & \thirdcell{19.13} & \secondcell{5.4} \\ \cline{2-7}
          & Histogram & 7.318 & 1006 & 5295 &  $1 \cdot 10^7$ & \firstcell{\textbf{0.1}} \\ \cline{2-7}
          & Sampling &2.464 &55.29 &276.1 & $4 \cdot 10^4$ &63 \\ \cline{2-7}
          &NeuroCard & 1.580 & 4.545 & 5.910 & \firstcell{\textbf{8.510}} & 673 \\ \cline{2-7}
          &DeepDB &\thirdcell{1.318} & \secondcell{2.500} & \secondcell{3.161} & 39.60 & 49 \\ \cline{2-7}
          &FLAT &\firstcell{\textbf{1.150}} & \firstcell{\textbf{1.819}} & \firstcell{\textbf{2.247}} & \secondcell{10.86} & 6.8 \\ \cline{2-7}
          &MSCN & 2.750 &19.70 &97.60 & 661.0 & \thirdcell{6.7} \\
			\thickhline
			\multirow{7}{*}{JOB-Comp} 
			&\textbf{\textit{BayesCard}} & \secondcell{1.271} & \secondcell{9.053} & \thirdcell{86.3} & \secondcell{$4 \cdot 10^4$} & \secondcell{6.2}\\ \cline{2-7}
			&Histogram & 15.78 & 7480 & $4\cdot10^4$ & $1\cdot10^8$ & \firstcell{\textbf{0.2}} \\\cline{2-7}
          &Sampling & 3.631 & 102.7 & 1374 & $8\cdot10^6$ & 101 \\ \cline{2-7}
          &NeuroCard &\thirdcell{1.538} & \thirdcell{9.506} & \secondcell{81.23} &  \thirdcell{$1 \cdot 10^5$} & 73\\ \cline{2-7}
          &DeepDB & 1.930 & 28.30 & 248.0 & $1 \cdot 10^5$ &55\\  \cline{2-7}
          &FLAT &\firstcell{\textbf{1.202}} & \firstcell{\textbf{6.495}} & \firstcell{\textbf{57.23}} & \firstcell{$\boldsymbol{1\cdot10^4}$} & 10.1\\ \cline{2-7}
          &MSCN & 4.961 &45.7 &447.0 & $1\cdot10^5$ & \thirdcell{6.6} \\
			\hline
	\end{tabular}}
	\vspace{-1em}
	\label{tab: exp-multi}
\end{table}

\begin{figure}[t]
  \centering
  \includegraphics[width=8.5cm]{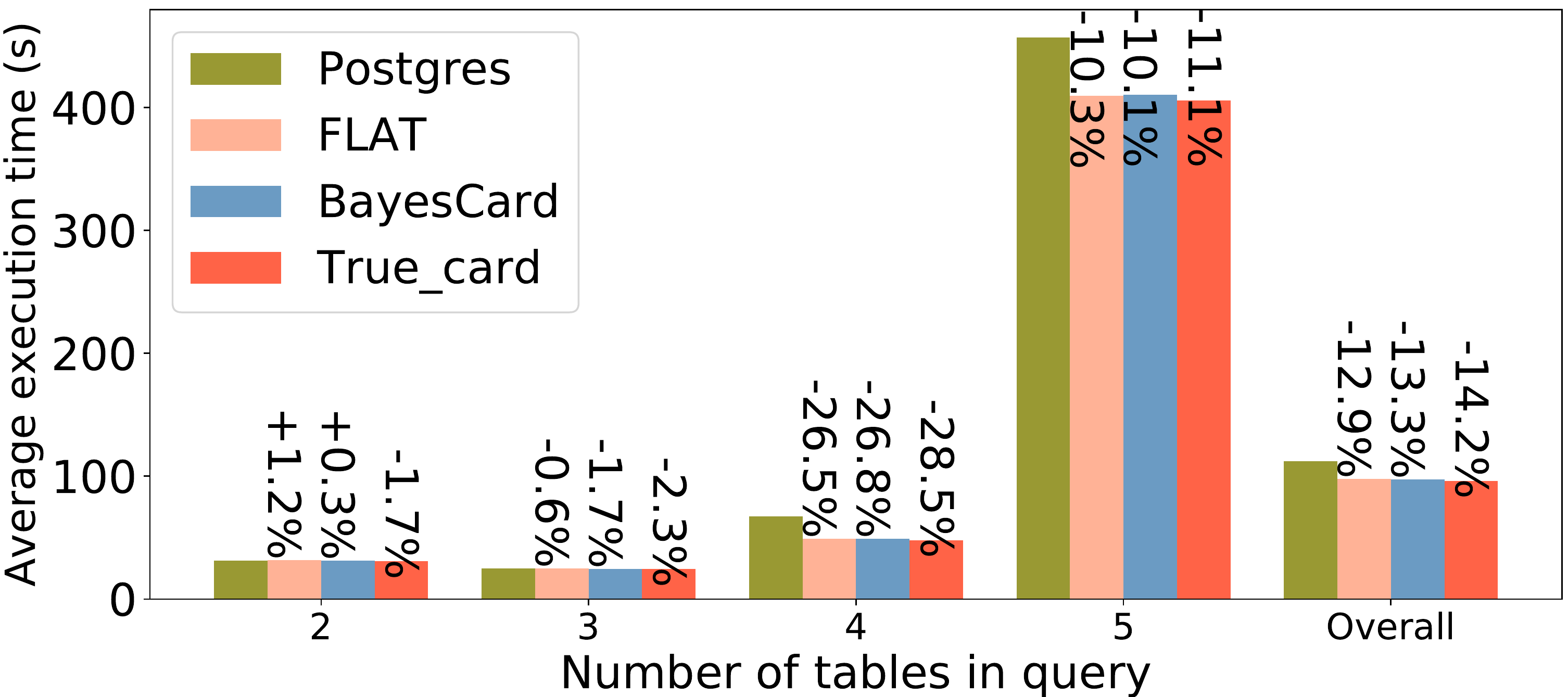}
  \vspace{-2.5em}
  \caption{ End-to-End evaluation of \textit{BayesCard}.}
  \vspace{-2.5em}
  \label{e2e}
\end{figure}

\noindent\underline{\textbf{End-to-End evaluation on PostgreSQL.}} 
Furthermore, we use the IMDB dataset to demonstrate \textit{BayesCard}'s behavior in terms of \emph{System} criteria. The four aspects of \emph{System} criteria are rather conceptual and hard to compare quantitatively in experiment, so we incorporate \textit{BayesCard} into a commercial DBMS, \emph{Postgres 9.6.6}, to show that it can improve the query optimization process of a real system. Specifically, we evaluate the end-to-end query processing time for JOB-light queries as shown in Figure~\ref{e2e}, and compare \textit{BayesCard} with the Postgres baseline, \textit{FLAT}, and optimal result derived by inserting the true cardinality during query optimization. We do not compare with other methods since \textit{FLAT} has established its SOTA performance in the same experiment, as reported in the original paper~\cite{zhu2020flat}. We observe that:

1) \textit{BayesCard} improves the Postgres baseline by $13.3\%$, suggesting that with more accurate \CE results, the query optimizer can generate better query plans with lower execution cost. 

2) The improvement of \textit{BayesCard} is very close to the method using true cardinality in query compiling (14.2\%). This verifies that the accuracy of \textit{BayesCard} is sufficient to generate high-quality query plans. Besides, even though \textit{BayesCard} has a slightly worse estimation accuracy, it still marginally outperforms \textit{FLAT}. Both methods produce similar execution plans and the marginal gain of \textit{BayesCard} over \textit{FLAT} mainly credits to its faster inference latency.

3) The improvement of \textit{BayesCard} and \textit{FLAT} becomes more significant on queries joining more tables because the execution plan for a query joining 2 or 3 is almost fixed. Whereas, for queries joining more tables, the inaccurate Postgres baseline results may lead to a sub-optimal query plan, while \textit{BayesCard} and \textit{FLAT} providing more accurate \CE results can find a better plan. This phenomenon has also observed and explained in~\cite{zhu2020flat, perron2019learned}.

\noindent\textbf{\textit{Summary:}}
The integration of \textit{BayesCard} into Postgres validates it as a practical counterpart of the \CE component in Postgres and also verifies that \textit{BayesCard} is a system-friendly \CE method.

\subsection{Comparing algorithms within \textit{BayesCard}} 
\label{sect6.4}
In this section, we compare different \textit{BayesCard}'s structure learning algorithms, perform ablation studies on the inference algorithms and summarize the take-home messages for using \textit{BayesCard}.

\begin{table}[t]
  \caption{Comparing different structure learning algorithms of \textit{BayesCard} on CENSUS.}
  \vspace{-1em}
	\resizebox{\columnwidth}{!}{
		\begin{tabular}{c|c|c|c|c|c}
			\hline
			\multirow{2}{*}{Algorithms} & 95\% & Infer. & Model & Train & Update \\
			& q-error & Time (s) & Size (mb) & Time (min) & Time (s) \\ \hline
			Exact & \textbf{1.24} & 16.5 & 43.7 & 298 & 1391\\ \hline
			Greedy & 1.88 & 2.45 & 2.53 & 62.1 & 442\\ \hline
			Chow-Liu &2.05 &\textbf{0.78} & \textbf{0.08} & \textbf{19.8} & \textbf{103} \\
			\hline
	\end{tabular}}
	\label{structlearn}
\end{table}

\smallskip
\noindent\underline{\textbf{Comparing structure learning algorithms.}} We report the estimation accuracy, inference latency without any proposed techniques, training time, model size, and update time on CENSUS dataset for various structure learning algorithms in Table~\ref{structlearn}. For \emph{exact} and \emph{greedy} algorithms, we incorporated the ``expert knowledge'' as described in Section~\ref{sect4.3}; otherwise, these algorithms become intractable and can not generate the BN's structure. We observe that with a more accurate structure learning algorithm (exact), the estimate accuracy has a significant improvement, but it sacrifices the other four dimensions to a great extent.
We did not report the result for DMV and IMDB datasets with a much fewer number of attributes because their data causal patterns are much simpler and different structure learning algorithms have similar performance.

\smallskip
\noindent\underline{\textbf{Ablation study of inference algorithms.}} We compare the novel inference optimizations of \textit{BayesCard} with the original variable elimination (VE) and belief propagation (BP) algorithms on a model learned with Chow-Liu tree algorithm on the CENSUS dataset, shown in Table~\ref{ablation}. We have the following observations: (1) the latency of original algorithms, VE and BP is unaffordable (780 ms per query) for practical systems; (2) the graph reduction (GR) and just-in-time compilation (JIT) optimization do not affect the estimation accuracy; (3) the GR and JIT alone improve the inference latency by 5 and 30 times respectively, and 325 times when combined for VE; (4) the progressive sampling algorithm (PS) produces 4 times larger estimation error but with significant improvement in latency. Worth noticing that the inference latency of PS and PS+GR can be much faster than VE+GR+JIT for \textit{BayesCard} with a complex structure (e.g. learned by exact structure learning algorithm).  

\smallskip
\noindent\underline{\textbf{Take-home messages for \textit{BayesCard} users.}} (1) The Chow-Liu tree structure learning algorithm can efficiently generate a compact model, which has improved inference latency and stable performance over other structure learning algorithms. The degrades in accuracy can be compensated using ``expert knowledge'' described in Section~\ref{sect4.3}. (2) The \emph{VE+GR+JIT} inference algorithm efficiently produces exact estimation for BNs with discrete attributes, which is debuggable, predictable, reproducible, and very friendly for system development. However, \emph{PS+GR} is a general approach that has guaranteed efficiency for any complex DAG-structured BN, and support continuous attributes with any distribution. (3) \textit{BayesCard} provides a general \CE framework for users to explore different trade-offs to suit their data and system settings.

\begin{table}[t]
  \caption{Ablation study of different inference algorithms of \textit{BayesCard} on CENSUS.} 
  \vspace{-1em}
	\resizebox{\columnwidth}{!}{
		\begin{tabular}{c|c|c|c|c|c|c|c}
			\hline
			Algorithms & VE & BP & VE+GR & VE+JIT & VE+GR+JIT &PS & PS+GR \\ \hline
			95\% q-error & \textbf{2.05} & \textbf{2.05} & \textbf{2.05} & \textbf{2.05} & \textbf{2.05} & 7.47 & 7.47 \\ \hline
			Latency (ms) & 780 & 685 & 190 & 21.9 & \textbf{2.4} & 8.8 & 3.5 \\
			\hline
	\end{tabular}}
	\vspace{-0.5em}
	\label{ablation}
\end{table}

\smallskip
\section{Related Work}
\label{sect8}

We will briefly revisit the existing \CE methods based on BN and the supervised \CE methods.

\smallskip
\noindent \underline{\textbf{BN-based methods}} have been explored decades ago for \CEend. Getoor et al.~\cite{2001SigmodGreedy} used a \textit{greedy} algorithm for BN structure learning, the variable elimination for probability inference, and referential integrity assumption for join estimation. Tzoumas et al.~\cite{tzoumas2011lightweight} learned an exact-structured BN and used belief propagation for inference. Halford et al.~\cite{dasfaa2019} adopted the Chow-Liu tree structure learning algorithm, the VE inference algorithm, and the uniformity assumption for join estimation. However, none of the practical DBMSes incorporates these methods due to their impractical structure learning process, intractable inference latency, or inaccurate estimation for join queries due to over-simplified assumptions.

\smallskip
\noindent \underline{\textbf{Supervised \CE methods}} use the feedback of past queries to train ML models, which maps the featurized query $Q$ to its actual cardinality. The first approach using neural networks on cardinality estimation was published for UDF predicates~\cite{5}. Later on, a regression-based model~\cite{7} and a semi-automatic alternative~\cite{8} were presented. Recently, supervised DL-based approaches, used multi-set convolutional network (\textit{MSCN})~\cite{MSCN}, tree-LSTM~\cite{sun2019end}, and lightweight XG-boost model~\cite{dutt2019selectivity} for \CEend. However, the supervised learning approaches have two major drawbacks as mentioned in ~\cite{deepDB}: (1) Their models neglect the data itself and are not robust to changes in query workload.
(2) Collecting the training data can be very expensive and training data has to be recollected when the workload changes.
Therefore, in general, query-driven supervised ML methods on cardinality estimation are not as flexible and accurate as data-driven unsupervised ML methods.

\smallskip
\section{Conclusion}
\label{sect8}
This paper proposes \textit{BayesCard}, the first framework that unifies the existing efforts on PPLs and BNs and optimizes them for \CE  in different data and system settings. \textit{BayesCard} revitalizes BNs with new equipments in model construction and probability inference, which make it a desirable \CE method satisfying the \emph{algorithm}, \emph{data} and \emph{system} criteria at the same time. Extensive experimental studies and end-to-end system deployment establish \textit{BayesCard}'s superiority over existing \CE methods. 

Furthermore, \textit{BayesCard} captures the underlying data causality, which benefits other data-related tasks. In future work, we plan to explore the possibility of using \textit{BayesCard} for other tasks, such as data cleaning, entity matching, and approximate query processing.

\clearpage
\newpage

\bibliographystyle{ACM-Reference-Format}
\bibliography{refs}

\newpage
\appendix
\section{Proof of Theorem 1:}
\label{app1}
\noindent\textit{\textbf{Theorem 1}. Given a BN and its defined DAG $G = (V, E)$, representing a table $T$ with attributes $V$ = \{$T_1, \cdots T_n$\}, and a query $Q=(T'_1=t'_1 \wedge \cdots \wedge T'_k=t'_k)$ where $T'_i \in V$. Let $G' = (V', E')$ be a sub-graph of $G$ where $V'$ equals $\bigcup_{1\leq i \leq k} Ancestor(T'_i)$, and $E'$ equals all edges in $E$ with both endpoints in $V'$. $Ancestor(T'_i)$ includes all parent nodes of $T'_i$ and all parents of parent node recursively. Then, performing \emph{VE} of BN on full graph G is equivalent to running \emph{VE} on reduced graph G'.}

\textbf{Proof of Theorem 1}: Given the probability query $Q$ on original graph $G$ and the reduced graph $G'$ defined above, we define $Q_V = \{T'_1, \cdots, T'_k\}$ and $V/Q_V = {T''_1, \cdots T''_{n-k}}$. In this proof, we will only show that running \emph{VE} on $G$ is equivalent to running \emph{VE} on $G'$. Then the proof for \emph{progressive sampling} naturally follows as it is directly approximating the computation of \emph{VE}.

First, recall that by law of total probability, we have the following Equation~\ref{equA1}.
\begin{align}
    P_T(T'_1=t'_1,& \cdots, T'_k=t'_k) = \sum_{t''_1 \in D(T''_1)} \cdots \sum_{t''_{n-k} \in D(T''_{n-k})} \nonumber \\
    &\Bigg[ \prod_{T'_i \in Q_V} P_T(T'_i = t'_i|Parents(T'_i)) * \nonumber \\
    & \;  \prod_{T''_i \in V/Q_V} P_T(T''_i = t''_i|Parents(T''_i)) \Bigg] 
    \label{equA1}
\end{align}
where $D(T''_i)$ denotes the domain of attribute $T''_i$ and $Parents(T''_i)$ denotes the parents of node $T''_i$ in graph $G$. For simplicity, here we refer to $Parents(T''_i)$ as ($T''_j = t''_j$, $\forall \ T''_j \in Parents(T'_i)$). The \emph{VE} algorithm are essentially computing Equation~\ref{equA1} by summing out one attribute from $V/Q_V$ at a time until all $T''_i \in V/Q_V$ are eliminated~\cite{PGM}.

Alternatively, we can derive the following Equation~\ref{equA2} by law of total probability and conditional independence assumption.
\begin{align}
    P_T&(T'_1=t'_1, \cdots, T'_k=t'_k) \nonumber \\
    &= \sum_{T''_i \in \bigcup Parents(T'_j)_{1\leq j \leq k}} \; \; \; \; \sum_{t''_i \in D(T''_i)}  \nonumber \\
    &\; \; \; \; \; \; \; \;  \; \Bigg[ P_T \Big(T'_1=t'_1, \cdots, T'_k=t'_k|\bigcup(Parents(T'_j)_{1\leq j \leq k}) \Big) * \nonumber \\
    &\; \; \; \; \; \; \; \;  \; \; P_T \Big(\bigcup Parents(T'_j)_{1\leq j \leq k} \Big) \Bigg] \nonumber \\
    & = \sum_{T''_i \in \bigcup Parents(T'_j)_{1\leq j \leq k}} \; \; \; \; \sum_{t''_i \in D(T''_i)}  \nonumber \\
    &\Bigg[ \prod_{T'_j \in Q_V} P_T(T'_j = t'_j|Parents(T'_j)) * P_T \Big(\bigcup Parents(T'_j)_{1\leq j \leq k} \Big) \Bigg]
    \label{equA2}
\end{align}

where $Parents(T''_i)$ denotes the parents of node $T''_i$ in graph $G$, which is the same as parents of node $T''_i$ in graph $G'$. By definition of reduced graph $G'$ where $V'$ = $\bigcup_{1\leq i \leq k} Ancestor(T'_i)$. $Ancestor(T'_i)$ includes all parent nodes of $T'_i$ and all parents of parent node recursively. Let $|V'| = n'$ and $V'/Q_V = T'''_1, \cdots, T'''_{n'-k}$. We can recursively write out $P_T \Big(\bigcup Parents(T'_j)_{1\leq j \leq k} \Big)$ using Equation~\ref{equA2} and result in Equation~\ref{equA3}.
\begin{align}
    P_T(T'_1=t'_1,& \cdots, T'_k=t'_k) = \sum_{t'''_1 \in D(T'''_1)} \cdots \sum_{t'''_{n'-k} \in D(T'''_{n-k})} \nonumber \\
    &\Bigg[ \prod_{T'_i \in Q_V} P_T(T'_i = t'_i|Parents(T'_i)) * \nonumber \\
    &\; \prod_{T'''_i \in V/Q_V} P_T(T'''_i = t'''_i|Parents(T'''_i)) \Bigg] 
    \label{equA3}
\end{align}

Equation~\ref{equA3} has the same form as Equation~\ref{equA1} with less attributes in the summation. Thus the \emph{VE} algorithm~\cite{PGM} can compute Equation~\ref{equA3} by eliminating one attribute from $V'/Q_V$ at a time. Thus running \emph{VE} on $G$ is equivalent to running \emph{VE} on $G'$.

\section{Computing the dependence level between tables}
\label{app2}
We use the randomized dependence coefficient (RDC)~\cite{rdc} as a measure of dependence level between two attributes. RDC is invariant with
respect to marginal distribution transformations and has low computational cost and it is widely used in many statistical methods~\cite{wu2020fspn, deepDB}. The complexity of RDC is roughly $O(n*log(n))$ where n is the sample size for the two attributes.

\begin{figure}[htp]
  \centering
  \includegraphics[width=8.5cm]{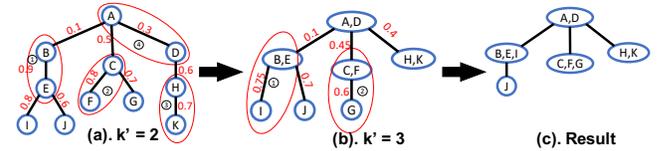}
  \caption{PRM Ensemble learning algorithm demonstration}
  \label{PRM_learn_app}
\end{figure}

\subsection{Calculating the pairwise RDC score between two tables}
\label{app21}
Recall Figure~\ref{PRM_learn_app}, we have a DB schema with 11 tables $A, \cdots, K$ and their join tables are defined as a tree $\mathbb{T}$ on the left image. In addition, we have unbiased samples $\mathbb{S}$ of the full outer join of all tables in $\mathbb{T}$ using the previously mentioned approach~\cite{Sampling}. Now consider $T, R \in$ {$A, \cdots, K$} as two random tables in this schema with attributes $T_1, \cdots, T_n$ and $R_1, \cdots, R_m$ respective. We can compute the pairwise RDC score between attributes $T_i$ and $R_j$, $RDC_{ij}$ based on $\mathbb{S}$, as described in~\cite{rdc}. Then we take the average as the level of dependence between $T$ and $R$ in the following Equation~\ref{equ:app1}.
\begin{equation}
    \sum_{1\leq i \leq n} \; \; \; \; \sum_{1\leq j \leq m} RDC_{i,j}/(n*m)
    \label{equ:app1}
\end{equation}

Thus, we can compute the dependence level matrix $M$ of size $11 \times 11$ with each entry specifying the dependence level between two tables in the schema. Then the edge weights of original $\mathbb{T}$ on the left image can be directly taken from $M$. The complexity of calculating $M$ is thus $O(m^2*|\mathbb{S}|*log(|\mathbb{S}|))$ where m is the total number attributes in all tables. 
 
\subsection{Calculating the pairwise RDC score between two set of tables}
During the PRM ensemble construction procedure, we sometimes need to calculate the dependence level between two sets of tables, such as the dependence level of {$A, D$} and {$H, K$} as in the right image of Figure~\ref{PRM_learn_app}. Similarly to the previous cases in Section~\ref{app21}, this value can be directly computed from $M$.

Take $Att(T)$ denotes the set of attributes in table T. Same as Equation~\ref{equ:app1}, the level of dependence between {$A, D$} and {$H, K$} is defined as Equation~\ref{equ:app2}.
\begin{align}
    &\sum_{ad \in Attr(\{A,D\})} \; \sum_{hk \in Attr(\{H,K\})} RDC_{ad,hk}/ \nonumber \\ 
    &  \; \;  \; \;  \; \;\Big(|Attr(A)+Attr(D)|*|Attr(H)+Attr(K)|\Big) \nonumber \\
    \; \; &= \Bigg(\sum_{a \in Attr(A)} \; \; \sum_{h  \in Attr(H)} RDC_{a,h} \nonumber \\
    & + \sum_{a \in Attr(A)} \; \; \sum_{k \in Attr(K)} RDC_{a,k} \nonumber \\
    &  + \sum_{d \in Attr(D)} \; \; \sum_{h \in Attr(H)} RDC_{d,h} \nonumber \\
    & + \sum_{d \in Attr(D)} \; \; \sum_{k \in Attr(K)} RDC_{d,k} \Bigg)\nonumber \\
    &\; \;  \; \;  \; \;/\Big(|Attr(A)+Attr(D)|*|Attr(H)+Attr(K)|\Big) \nonumber \\
    \; \; &= \Big( M[A, H] * |Attr(A)| * |Attr(H)| \\
    & + M[A, K] * |Attr(A)| * |Attr(K)| \nonumber \\
    & + M[D, H] * |Attr(D)| * |Attr(H)| \nonumber \\
    & +  M[D, K] * |Attr(D)| * |Attr(K)| \Big) \nonumber \\
    &\; \;  \; \;  \; \;/\Big(|Attr(A)+Attr(D)|*|Attr(H)+Attr(K)|\Big)
    \label{equ:app2}
\end{align}

Thus the weight of the edge can be updated quickly knowing the pre-computed $M$ and the number of attributes in each table.

\section{Additional experimental results}

The addition experiments on SYNTHETIC comparing \textit{BayesCard} with other methods are reported in Table~\ref{tabapp1} and Table~\ref{tabapp2}. Please note that we do not fine-tune the hyper-parameters of the DL-based methods since the training time on all the datasets take so long that we can not afford to explore different hyper-parameters. But we believe that the experimental results are enough to show the insights. On the other hand, \textit{BayesCard} do not have hyper-parameters to fine-tune, which is another advantage of our methods. We summarize our observations as follows.

\noindent\textbf{\text{Distribution (s):}}
For the inference latency and model size, we find that 
increasing the Pareto distribution skewness would degrade the performance of \textit{DeepDB} and \textit{FLAT}, but has little affect on all other methods. This is because \textit{DeepDB} and \textit{FLAT} tend to generate larger models on more complex data. As a result, their training time also improves w.r.t.~the skewness level.
The training time of \textit{Naru} and \textit{MSCN} also grows w.r.t.~skewness level, as their underlying DNNs need more time to model the complex distributions.

\noindent\textbf{\text{Correlation (c):}}
For the inference latency and model size, the increase of $c$ has negative impact on \textit{DeepDB} and \textit{FLAT}, as their models become larger on more correlated data. However, \textit{FLAT} behaves well on very highly correlated data since it can split them with other attributes to reduce the model size. The impact of $c$ on \textit{BayesCard} is mild, whose model size is still affordable.
For the training time, the increase of $c$ has impact on algorithms except \textit{Histogram} and \textit{Sampling}.
This is also reasonable as they need more time to model the complex distribution. Note that, for each setting of $c$, the training time of our \textit{BayesCard} is still much less than them.

\noindent\textbf{\text{Domain (d):}} 
For the inference latency, model size and training time, the increase of $d$ has significant impact on \textit{Naru}, \textit{DeepDB} and \textit{FLAT}. This is increasing the number of attributes would increases the data complexity exponentially, so they need more neurons or nodes to model the distribution. Whereas, the impact on our \textit{BayesCard} is much mild.

\noindent\textbf{\textit{Scale (n):}}
Similar to domain size, increasing the number of attributes also increases the data complexity exponentially, and thus we expect to see an increase in latency, model size and traininbg time for almost all methods. However, in comparison with \textit{Naru}, \textit{DeepDB}, \textit{FLAT} and \textit{MSCN}, the impact on \textit{BayesCard} is not very significant.

\noindent\textit{
\textbf{{Summary:}}
In comparison with DL-based \CE methods, our  \textit{BayesCard} attains very stable and robust performance in terms of inference latency, model size and training time.}

\begin{table*}[t]
    \centering
    \caption{Stability performance of different \CE method w.r.t. changes in data distribution skewness and correlation.}
    \vspace{-1em}
    \scalebox{0.8}{
    \begin{tabular}{|c|c|c|c|c|c|c|c||c|c|c|c|c|c|}
    \hline
     \CE & Algorithm & \multicolumn{6}{c|}{Distribution Skewness (s)} & \multicolumn{6}{c|}{Attribute Correlation (c)} 
           \\
     Methods & Criteria & \multicolumn{6}{c|}{c=0.4, d=100, n=10} & \multicolumn{6}{c|}{s=1.0, d=100, n=10}
           \\\cline{3-14}
     & & s=0 &s=0.3 &s=0.6 &s=1.0 &s=1.5 & s=2.0 &c=0 &c=0.2 &c=0.4 &c=0.6 &c=0.8&c=1.0 \\ \thickhline
     \multirow{4}{*}{\textbf{\textit{BayesCard}}} 
     & Accuracy (95\% q-error) & 1.06 &\textbf{1.09} &1.25 &\textbf{1.49} &2.39 & 2.28 & \textbf{1.32} &1.28 & 1.48 &2.13 & \textbf{1.49} & \textbf{1.00} \\ \cline{2-14}
     & Latency (ms) & 3.5  & 2.8 & 3.7 & 2.4 & 2.2 & 3.0 & \textbf{0.1} & 2.9 &2.4 & 1.6 &3.3 & 2.1 \\ \cline{2-14}
     & Model size (kb) & 534 & 530 & 538 & 529 &403 & 514 & 9.5 & 525 & 508 & 478 & 605 & 396 \\ \cline{2-14}
     & Training time (s) & 17.5 & 18.2 & 17.3 & 19.2 & 16.8 & 14.4 & 5.2 & 19.2 &17.3 & 37.9 & 21.3 & 45.6 \\ \thickhline
     \multirow{4}{*}{Histogram}
     & Accuracy (95\% q-error) & 136 & 112 &195 &240 &219 & 277 & \textbf{1.32}  &73.2 & 240 &1403 & $2\cdot 10^4$ & $9\cdot 10^4$ \\ \cline{2-14}
     & Latency (ms) & \textbf{0.1} & \textbf{0.1} & \textbf{0.1} &\textbf{0.1} &\textbf{0.1}& \textbf{0.1}  &\textbf{0.1} &\textbf{0.1}  &\textbf{0.1} &\textbf{0.1}  &\textbf{0.1}  &\textbf{0.1}  \\ \cline{2-14}
     & Model size (kb) & \textbf{11.5} & \textbf{9.7} & \textbf{8.2} & \textbf{8.8} & \textbf{7.3} & \textbf{7.5} & \textbf{9.5} & \textbf{8.3} & \textbf{8.8} & \textbf{9.7} & \textbf{8.4} & 9.1 \\ \cline{2-14}
     & Training time (s) & \textbf{3.1} & \textbf{3.0} & \textbf{3.1} & \textbf{4.4} & \textbf{3.9} & \textbf{4.1} & \textbf{3.0} & \textbf{2.9} & \textbf{3.4} & \textbf{3.0} & \textbf{3.3} &\textbf{3.1} \\ \thickhline
     \multirow{4}{*}{Sampling}
     & Accuracy (95\% q-error) & \textbf{1.03} & 1.67 & 2.87 & 2.21 & 55.9 & 142.1 &1.78 &1.62 & 2.21 & 4.07 & 2.10 & 1.03 \\ \cline{2-14}
     & Latency (ms) & 52 & 54 & 52 & 60 & 52 & 49 & 51 & 53 & 61 & 47 & 52 & 52\\ \cline{2-14}
     & Model size (kb) & - & - & - & - & - & - & - &- &- &- &- & - \\ \cline{2-14}
     & Training time (s) & - & - & - & - & - & - & - &- &- &- &- & - \\ \thickhline
     \multirow{4}{*}{Naru}
     & Accuracy (95\% q-error) & \textbf{1.03} & 1.22 & 1.78 & 3.62 & 28.8 & 21.4 &1.76 & \textbf{1.23} & 3.62 & \textbf{2.09} & 1.71 & 1.10 \\ \cline{2-14}
     & Latency (ms) & 58 & 67 & 58 & 62 & 60 & 66 & 73 & 65 & 62 & 61 & 66 & 59 \\ \cline{2-14}
     & Model size (kb) & 3670 & 3670 & 3670 &3670 &3670 & 3670 &3670 &3670 &3670 &3670 & 3670& 3670\\ \cline{2-14}
     & Training time (s) & 4460 & 4910 & 4879 & 5503 & 5908 & 6371 & 1702 & 4702 & 5503 & 9915 &4962& 1308 \\ \thickhline
     \multirow{4}{*}{DeepDB}
     & Accuracy (95\% q-error) & 4.51 & 4.89 & 15.1 & 14.0 & 14.2 & 19.0 & 1.32 & 3.58 & 15.1 & 117 & 663 & 108  \\ \cline{2-14}
     & Latency (ms) & 4.4 & 4.6 & 7.5 & 7.3 & 5.8 & 9.2 & \textbf{0.1} & \textbf{4.3} &7.3 & 10.6 & 16.4 & 13.2\\ \cline{2-14}
     & Model size (kb) & 701 & 597 & 834 & 1107 & 1104 & 1315 &9.5 & 570 & 1198 & 1864 & 5532 & 1907 \\ \cline{2-14}
     & Training time (s) & 131 & 133 & 197 & 247 & 271 & 280 &5.5 & 131 &244.2 & 421 & 2570 & 830 \\ \thickhline
     \multirow{4}{*}{FLAT}
     & Accuracy (95\% q-error) & 1.06 & 1.15 & \textbf{1.23} & 1.76 & \textbf{2.25} & \textbf{2.11} &1.32 & 1.27 & 1.76 & 2.11 & 1.73 &\textbf{1.00} \\ \cline{2-14}
     & Latency (ms) & 0.6 & 0.9 & 0.6 & 0.5 &1.5 &1.7 & \textbf{0.1} & 0.7 &0.5 & 4.1 &17.8 & 0.2 \\ \cline{2-14}
     & Model size (kb) & 76 & 101 & 80 & 75 &430 & 580 &9.5 &103 &75 & 1201 & 1889 & \textbf{4.7} \\ \cline{2-14}
     & Training time (s) & 91 & 93 & 127 & 142 & 240 & 253 &5.5 & 133 &244.2 & 629 & 1370 & 17.0 \\ \thickhline
     \multirow{4}{*}{MSCN}
     & Accuracy (95\% q-error) & 55.1 & 160 &105 & 94 & 129 & 340 & 51.3 & 54.0 &94 &145 & 544 &620 \\ \cline{2-14}
     & Latency (ms) & 1.1 &1.1 &1.0 &1.1 &1.3 &1.0 &1.0 &1.4 &1.2 &1.0 & 1.3 & 1.1 \\ \cline{2-14}
     & Model size (kb) & 3200 &3200 &3200 &3200 &3200 & 3200 &3200 &3200 &3200 &3200 &3200&3200 \\ \cline{2-14}
     & Training time (s) & 1203 & 1208 & 959 & 1430 &871 & 1770 & 922 &935 &1432 &955 &1831 & 880 \\ \thickhline
    \end{tabular}
    \label{tabapp1}}
\end{table*}

\begin{table*}[t]
    \centering
    \caption{Scalability performance of different \CE method w.r.t. changes in data domain size and number of attributes.}
    \vspace{-1em}
    \scalebox{0.8}{
    \begin{tabular}{|c|c|c|c|c|c|c|c||c|c|c|c|c|c|}
    \hline
     \CE & Algorithm & \multicolumn{6}{c|}{Domain Size (d)} &  \multicolumn{6}{c|}{Number of Attributes (n)}
           \\
     Methods & Criteria & \multicolumn{6}{c|}{s=1.0, c=0.4, n=10} & \multicolumn{6}{c|}{s=1.0, c=0.4, d=100} 
           \\\cline{3-14}
     & &d=10 &d=100 &d=500 &d=1000 &d=5000 &d=10000 &n=2 &n=5 &n=10 &n=50 &n=100&n=200 \\ \thickhline
     \multirow{4}{*}{\textbf{\textit{BayesCard}}} 
     & Accuracy (95\% q-error) & 1.29 &\textbf{1.49} &\textbf{1.05} & \textbf{1.04} & 25.3 & 49.1 & 1.04 &1.12 &\textbf{1.49} & \textbf{2.58} & \textbf{1.97} & \textbf{3.02} \\ \cline{2-14}
     & Latency (ms) & 1.0 & 2.4 & 2.8 & 11.5 & 2.2 & 3.6 & 0.4 & 1.5 & 2.4 & 4.7 &11.3 &15.1 \\ \cline{2-14}
     & Model size (kb) & 19.3 &542 &982 & 1280 & 168 & 417 & 64.2 & 236 &542 &2820 &5400 & $1\cdot10^4$ \\ \cline{2-14}
     & Training time (s) & 11.1 &14.9 &33.4 & 48.7 & \textbf{7.3} & \textbf{6.5} & 2.01 & 4.73 & 14.5 & 113 & 576 & 1907 \\ \thickhline
     \multirow{4}{*}{Histogram}
     & Accuracy (95\% q-error) & 15.2 &240 & 498 & 1066 & $2\cdot10^4$ & $6\cdot10^4$ & 19.3 & 103 & 240 & $3\cdot10^4$ & $1\cdot10^5$ & $9\cdot10^5$ \\ \cline{2-14}
     & Latency (ms) & \textbf{0.1} &\textbf{0.1} &\textbf{0.2} &\textbf{0.4} & \textbf{0.8} & \textbf{0.1} & \textbf{0.1} & \textbf{0.1} &\textbf{0.1} & \textbf{0.5} & \textbf{0.4} & \textbf{0.7} \\ \cline{2-14}
     & Model size (kb) & \textbf{0.8} & \textbf{9.8} & \textbf{14.0} & \textbf{12.8} &\textbf{13.2} & \textbf{14.3} & \textbf{1.5} & \textbf{4.0} & \textbf{10.1} & \textbf{48.8} & \textbf{173} & \textbf{308} \\ \cline{2-14}
     & Training time (s) & \textbf{1.1} & \textbf{3.0} & \textbf{4.3} & \textbf{6.7} &11.5 & 15.0 & \textbf{0.8} & \textbf{1.6} & \textbf{3.1} & \textbf{16.4} & \textbf{44.8} & \textbf{90.7} \\ \thickhline
     \multirow{4}{*}{Sampling}
     & Accuracy (95\% q-error) & 1.95 &2.21 & 2.09 & 4.93 & \textbf{21.2} & 57.9 & 1.04 & 1.26 &2.21 & 342 & $2\cdot 10^4$ & $3\cdot10^4$ \\ \cline{2-14}
     & Latency (ms) & 59 & 55 & 60 & 63 & 57 & 59 & 45 & 60 & 77 & 109 & 123 & 135 \\ \cline{2-14}
     & Model size (kb) & - & - & - & - & - & - & - &- &- &- &- & - \\ \cline{2-14}
     & Training time (s) & - & - & - & - & - & - & - &- &- &- &- & - \\ \thickhline
     \multirow{4}{*}{Naru}
     & Accuracy (95\% q-error) & \textbf{1.02} &3.62 & 7.30 & 89.4 & 292 & 1783 & 1.39 & \textbf{1.09} & 3.62 & 121.0 & 475 & 1098 \\ \cline{2-14}
     & Latency (ms) &32 &62 &81 &115 &154 & 260 & 33 & 37 & 62 & 225 &473 & 726 \\ \cline{2-14}
     & Model size (kb) & 3140 &3670 & 6135 & 8747 & $2\cdot10^4$ & $4\cdot10^4$ & 2447 &3050 &3670 & 6673 & 8010 &  $1\cdot10^4$ \\ \cline{2-14}
     & Training time (s) & 1032 &5503 & 5607 & 6489 & 5980 & $1\cdot10^4$ &1157 & 4201 &5503 &6930 & $1\cdot10^4$ &$2\cdot10^4$ \\ \thickhline
     \multirow{4}{*}{DeepDB}
     & Accuracy (95\% q-error) & 1.08 &10.0 & 15.1 & 107 & 61 & 213 & 1.04 & 1.17 & 15.1 & 257 & 1490 & $1\cdot10^4$ \\ \cline{2-14}
     & Latency (ms) & 1.3 &7.5 &8.8 & 11.9 & 10.7 & 19.0 & 0.6 & 1.4 & 7.5 & 25.4 & 67.1 & 109 \\ \cline{2-14}
     & Model size (kb) & 29.5 & 834 & 1234 & 1910 & 1781 & 3974 & 35.0 & 129 &834 & 6710 & $3\cdot 10^4$ &$9\cdot 10^4$ \\ \cline{2-14}
     & Training time (s) & 25 &197 & 769 & 2310 & 4155 & $1\cdot 10^4$ & 21 & 65 &197 &3698 & 8930 & $2\cdot 10^4$ \\ \thickhline
     \multirow{4}{*}{FLAT}
     & Accuracy (95\% q-error) & 1.08 &1.76 & 1.35 & 1.17 & 27.6 & \textbf{44.0} &\textbf{1.02} & \textbf{1.09} &1.76 & 255 & 2015 &$1\cdot10^4$ \\ \cline{2-14}
     & Latency (ms) & 0.5 & 0.6 & 1.5 & 18.0 & 15.9 & 49.7 &0.4 &0.5 &0.5 & 25.9 & 66.0 & 110 \\ \cline{2-14}
     & Model size (kb) & 16.1 &75.3 & 310 & 2701 & 1980 & 5732 & 15.0 & 49.9 &75.3 &6908 & $3\cdot 10^4$ &$9\cdot 10^4$ \\ \cline{2-14}
     & Training time (s) & 15.5 &142 & 198 & 2670 & 1535 & 9721 &9.7 &48.6 &142 &4017 & $1\cdot 10^4$ & $2\cdot 10^4$ \\ \thickhline
     \multirow{4}{*}{MSCN}
     & Accuracy (95\% q-error) & 53.0 & 94.4 & 106 & 188 & 173 & 290 & 18.8 & 40.5 &94.4 & 1783 & 8084 & $3\cdot 10^4$ \\ \cline{2-14}
     & Latency (ms) & 0.9 &1.1 &1.1 & 1.1 &1.0 & 1.2 & 0.4 &0.9 &1.1 &1.3 & 1.8 & 2.0 \\ \cline{2-14}
     & Model size (kb) & 3200 &3200 &3200 &3200 &3200 & 3200 & 2430 & 2871 &3200 & 3328 & 3609 & 3827 \\ \cline{2-14}
     & Training time (s) & 1179 &1430 & 1329 & 1398 & 1530 & 1230 & 719 & 821 & 1430 & 1600 & 2031 & 2299 \\ \thickhline
    \end{tabular}
    \label{tabapp2}}
\end{table*}

\end{document}